\documentclass[aps,prx,twocolumn,natbib,showpacs,floatfix,superscriptaddress]{revtex4-1}

\usepackage{graphicx}
\usepackage{amsmath,amssymb,bbold,bm,color}
\usepackage{float}
\usepackage{epstopdf}
\usepackage{hyperref}
\usepackage{color}
\usepackage{enumerate}
\usepackage{wasysym}
\usepackage{url}
\usepackage{dsfont}
\usepackage{braket}
\hypersetup{colorlinks=true,linkcolor=blue,citecolor=blue,urlcolor=blue}
\graphicspath{{figures/}{../figures/}}

\newcommand{\bk}{{\bm k}}

\newcommand{\br}{{\bm r}}
\newcommand{\bv}{{\bm v}}

\newcommand{\bB}{{\bm B}}

\newcommand{\bA}{{\bm A}}
\newcommand{\ba}{{\bm a}}
\newcommand{\bS}{{\bm S}}
\newcommand{\bL}{{\bm L}}
\newcommand{\bJ}{{\bm J}}

\newcommand{\bj}{{\bm j}}
\newcommand{\bsig}{{\bm \sigma}}

\newcommand{\cC}{{\cal C}}

\newcommand{\cT}{{\cal T}}

\newcommand{\cH}{{\cal H}}

\newcommand{\bee}{\begin{equation}}
\newcommand{\ee}{\end{equation}}

\def\sgn{\mathop{\rm sgn}\nolimits}

\begin{document}

\title{Probing topological degeneracy on a torus using superconducting
  altermagnets }

\author{Tsz Fung Heung}

\affiliation{Department of Physics, the Hong Kong University of Science and Technology, Clear
  Water Bay, Hong Kong, China}
\affiliation{Department of Physics and Astronomy, and Quantum Matter
  Institute, University of British Columbia, Vancouver, BC, Canada V6T 1Z1}

\author{Marcel Franz}

\affiliation{Department of Physics and Astronomy, and Quantum Matter
  Institute, University of British Columbia, Vancouver, BC, Canada V6T 1Z1}

\begin{abstract}
 The notion of topological order (TO) can
  be defined through the characteristic ground state degeneracy of a system placed  on a manifold with non-zero genus $g$, such as a torus.
  This ground state degeneracy has
served as a key tool for identifying TOs in theoretical calculations but it
has never been possible to probe experimentally because fabricating a
device in the requisite toroidal geometry is generally not
feasible.  Here we discuss a practical method that can be used to overcome this
difficulty in a class of topologically ordered systems that consist of a
TO and its time reversal conjugate $\overline{\rm TO}$.  The key
insight is that a system
possessing such ${\rm  TO}\otimes\overline{\rm TO}$ order fabricated on
an annulus behaves effectively as TO on a torus, provided that one supplies a
symmetry-breaking perturbation that gaps out the edge modes. We illustrate this general
principle using a specific example of a spin-polarized $p_x\pm ip_y$
chiral superconductor which is closely related to the Moore-Read Pfaffian
fractional quantum Hall state. Specifically, we introduce a simple
model with altermagnetic normal state which, in the presence of
an attractive interaction, hosts a helical
$(p_x-ip_y)^\uparrow\otimes(p_x+ip_y)^\downarrow$  superconducting ground
state. We demonstrate that when placed on an annulus with the
appropriate symmetry-breaking edge perturbation this planar two-dimensional
system, remarkably, exhibits the same pattern of ground state
degeneracy as a $p_x+ ip_y$ superconductor on a
torus. We discuss broader implications of this behavior and ways it
can be tested experimentally.   

\end{abstract}

\date{\today}
\maketitle

\section{Introduction}

When a 2D system with topological order is placed on a torus (or a
surface with higher-genus $g$), its ground state exhibits a degeneracy that
depends on $g$ while the member states of the degenerate ground manifold cannot be
distinguished by any local measurement
\cite{Haldane1985,Wen1990,Wen1991}. This key feature  forms 
a basis of early proposals to exploit TO systems in fault-tolerant
quantum computation. For practical
reasons, however, it is generally 
not possible to prepare 2D quantum systems in the toroidal geometry and
hence the ground state degeneracy in its basic form has never been
experimentally tested, even for well-established TOs occurring  in
fractional quantum Hall (FQH) phases. Instead, experimental work to-date \cite{Manfra2020,Heiblum2023} has focused
on probing either the gapless edge modes in more conventional planar
geometries with edges, guaranteed to be present by
the bulk-boundary correspondence, or the excitations in the
TO bulk, whose fractional charge and exchange statistics
\cite{Leinaas1977,Wilczek1982,Wen1991b,Greiter2024} (possibly
non-Abelian) are fundamentally related to the ground state degeneracy
on torus \cite{Verlinde1988,Einarsson1990,Oshikawa2006, Oshikawa2007}.   

For the same reason, attempts to employ TO systems in quantum
information storage and processing schemes have focused largely on directly manipulating
and probing its non-Abelian anyon excitations  \cite{Kitaev2003,Nayak2008}, often utilizing gapless edge
modes in the process. The presence of gapless excitations, however,
can make such devices sensitive to noise and decoherence, thus
subverting the promise of intrinsic fault tolerance. Clearly, to
realize the full potential of fault tolerance inherent to topological orders, it would
be preferable to use the ground state degeneracy in a fully gapped TO 
system for this purpose.
\begin{figure}[t]
  \includegraphics[width=8.1cm]{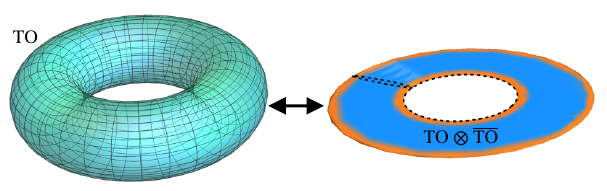}
  \caption{Equivalence of TO on the torus and ${\rm
      TO}\otimes\overline{\rm TO}$ on an annulus with edge modes
    gapped through a symmetry-breaking perturbation, marked here in orange. }
  \label{fig1}
\end{figure}

Recent work on ``Cheshire qudits'' \cite{Wen2024} advanced a visionary proposal
along these lines based on an even-denominator fractional quantum spin
Hall (FQSH) state that may be present in twisted homobilayer MoTe$_2$
according to the recent experimental report \cite{Mak2024}. One candidate state
consistent with the reported transport signatures is a helical
non-Abelian phase composed of a Moore-Read Pfaffian state
\cite{Moore1991} for spin-up
electrons and its time-reversal
($\cT$) conjugate for spin-down electrons \cite{Reddy2024,Ahn2024}, or ${\rm
  Pf}^\uparrow\otimes\overline {\rm Pf}^\downarrow$ for short. This
FQSH phase exhibits a pair of gapless helical edge modes protected by a
combination of time-reversal, charge conservation and spin/valley
symmetries. As noted in Ref.\ \cite{Wen2024}, a symmetry-breaking
perturbation at the edge can gap out 
the edge modes. This allows electrons near the edges to
tunnel between the  ${\rm  Pf}^\uparrow$ and $\overline {\rm
  Pf}^\downarrow$ states thus in effect  transforming a planar geometry with open
edges to a singly connected surface supporting the TO. In this way, a
planar flake of twisted homobilayer MoTe$_2$ with gapped edges can be
thought of as having a topology of a sphere with ${\rm  Pf}^\uparrow$
forming its top surface, $\overline {\rm Pf}^\downarrow$ its bottom
surface and the symmetry-breaking perturbation gluing the two halves
together. Similarly, a flake with a hole punched through its middle can be
thought of as a torus pictured in Fig.\ \ref{fig1}.
\begin{figure}[t]
  \includegraphics[width=8.5cm]{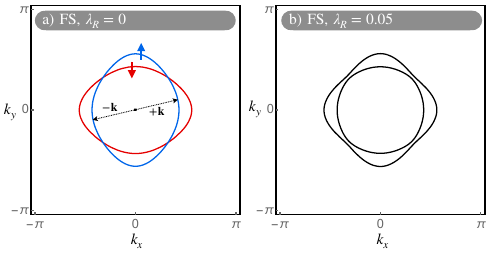}
  \caption{Spin split Fermi surfaces of the altermagnetic metal
    described by Hamiltonian Eqs.\ (\ref{h0},\ref{h1}) for $\eta=0.2$ and
  $\mu=-2.1$, with (b)  and without (a)  Rashba SOC. }
  \label{fig2}
\end{figure}

In Ref.\ \cite{Wen2024}  this equivalence was argued at the level of topological field
theory. The purpose of this work is to perform a study in the
framework of a microscopic model which can offer a complementary and
potentially deeper understanding of the physics that is involved. To
this end we employ the well-known equivalence between the Moore-Read
Pfaffian state and the weak-pairing state of fermions realizing a
spin-polarized $p_x\pm ip_y$ superconductor, or $p_\pm$ for
short. According to Read and Green \cite{Read2000} the many-body wavefunctions of the
two systems are asymptotically the same. As a result, they
also share many key physical properties, including the edge states,
quasihole exchange statistics and toroidal ground state degeneracy.

To
simulate a toroidal geometry using the equivalence principle depicted in Fig.\
\ref{fig1} we construct a simple model of 2D electrons whose ground
state in the presence of a weak attractive interaction can be described
as a helical $(p_x-ip_y)^\uparrow\otimes(p_x+ip_y)^\downarrow$
superconductor, abbreviated in the following as
$p_-^\uparrow\otimes p_+^\downarrow$.  The construction is based on a normal state that can be
characterized as an altermagnetic metal
\cite{Hayami2020,Smejkal2022a,Smejkal2022b,Mazin2022} with a pair of spin-split
Fermi surfaces related by a $C_4$ rotation and time reversal $\cT$,
pictured in Fig.\ \ref{fig2}(a). The key observation is that such a
spin-split Fermi surface cannot  support the conventional spin singlet
pairing; instead the leading instability is 
toward an odd-parity, $\cT$ broken $p_x\pm ip_y$ chiral phase in each
spin channel. We emphasize that this unconventional pairing state does
not require any exotic microscopic mechanism or unusual form of the
attractive interaction; instead it 
is a generic property of the spin-split fermi surface characteristic
of an altermagnet and will form through the conventional BCS mechanism in
the presence of  e.g.\ phonon-mediated attraction.

Furthermore, we show that a weak Rashba
spin-orbit coupling (SOC), acting either in the bulk or near the
edges, selects a state with opposite chiralities for two spin species
and hence stabilizes the $p_-^\uparrow\otimes
p_+^\downarrow$ helical phase. Intrinsic or proximity-generated spin-singlet SC
order at the edge  has the same effect. Importantly, 
 these same perturbations also gap out the helical edge
modes that would otherwise exist in the pure
$p_-^\uparrow\otimes p_+^\downarrow$ phase. Hence, we argue that
such a 
simple minimal model possesses all the ingredients that are necessary
to test the annulus/torus correspondence. Also, since several metallic 
altermagnet candidate materials have recently been experimentally
confirmed \cite{Liu2024,Krempasky2024,Lee2024,Fedchenko2024,Reimers2024},
altermagnets with SC order 
may offer a platform to probe the topological degeneracy in a
simple setting, in addition to providing insights into the physics of the
helical Moore-Read Pfaffian state that may be present in twisted
homobilayer MoTe$_2$.

A BCS-type Hamiltonian describing superconducting states conserves
electron number parity but not the electron number
itself. Accordingly, the eigenstates of such Hamiltonians can be
classified as parity even and parity odd. The chiral $p_x+ip_y$
superconductor exhibits four-fold ground state 
degeneracy when placed on a torus. Importantly, three of those states
belong to the even parity sector and one to the odd parity
sector \cite{Read2000}. This subtle feature is deeply rooted in the
non-Abelian exchange statistics of vortices in the chiral $p$-wave
state \cite{Ivanov2001} and distinguishes it from an ordinary $s$-wave superconductor
\cite{Oshikawa2006, Oshikawa2007}. When placed on a  
torus, the latter also exhibits four-fold ground-state degeneracy
\cite{Hansson2004}  but
in this case all four ground states belong to the even parity
sector. In the following we shall review the origin of this ground
state degeneracy and its robustness against disorder. We then
demonstrate the even/odd parity effect in the
ground state manifold  of the chiral $p_\pm$ state on the torus and the
helical  $p_-^\uparrow\otimes p_+^\downarrow$ on an annulus with
gapped boundaries, thus establishing their equivalence. In closing we
discuss broader significance of these results and ways these effects
can be probed experimentally.

\section{Microscopic model}
\label{sec:model}

\subsection{Altermagnetic normal metal}

As the starting point we consider a minimal model of a `$d$-wave' altermagnetic
metal \cite{Smejkal2022a,Smejkal2022b}  in two dimensions defined by the Hamiltonian
$\cH_0=\sum_\bk\psi_\bk^\dagger h_0(\bk)\psi_\bk$ with
$\psi_\bk=(c_{\bk\uparrow}, c_{\bk\downarrow})^T$ and
\begin{equation}\label{h0}
h_0(\bk)=-2t(\cos{k_x}+\cos{k_y})-2\eta\sigma_z(\cos{k_x}-\cos{k_y}).
\end{equation}
Here $\sigma_\mu$ are Pauli matrices in spin space, $t$ denotes the nearest
neighbor hopping amplitude on the square lattice and $\eta$ controls
the altermagnetic band spitting. The latter breaks time-reversal
symmetry $\cT$ but the model remains invariant under combined $C_4$
rotation and $\cT$, a hallmark feature of altermagnets which
guarantees vanishing total magnetization. The spin-split Fermi surfaces following from Eq.\
\eqref{h0} are depicted in Fig.\ \ref{fig2}(a). This type of
non-relativistic spin splitting has recently been experimentally
observed in MnTe \cite{Krempasky2024,Lee2024}, RuO$_2$
\cite{Fedchenko2024} and CrSb thin films \cite{Reimers2024}. In
addition, recent theoretical work
\cite{Naka2019,Yuan2020,Mazin2021,Spaldin2022,Guo2023,Ostanin2024}
identified several families of other materials as candidate altermagnets.

When placed on a substrate the inversion symmetry of $\cH_0$ will
generically be broken and a Rashba SOC term
\begin{equation}\label{h1}
h_R(\bk)=2\lambda_R(\sigma_x\sin{k_y}-\sigma_y\sin{k_x})
\end{equation}
becomes symmetry-allowed. For $\lambda_R$ 
non-zero Fermi crossings along the Brillouin zone diagonals are
split, Fig.\ \ref{fig2}(b)  as the Rashba term causes the spins to rotate into the plane in
their vicinity. In the following, we will consider superconducting
instabilities of $\cH_0$, and, when specified assume weak SOC with
magnitude small compared to the SC gap. We will take
$t=1$ and express all energies in units of $t$.
\begin{figure}[t]
  \includegraphics[width=8.5cm]{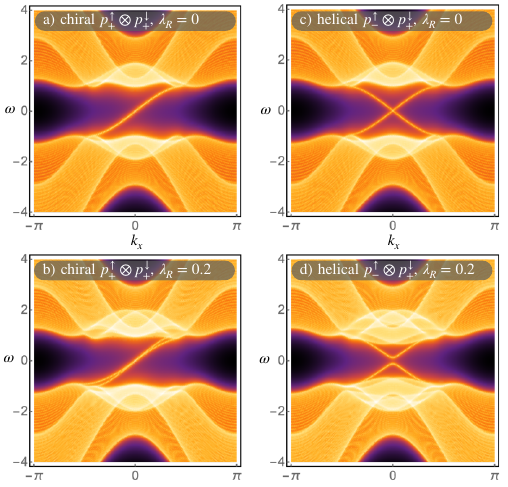}
  \caption{Edge states in chiral and helical phases. Spectral function $A_y(\omega,k_x)=-2{\rm
    Im}[\omega+i\delta-H(k_x)]_{yy}^{-1}$ for the superconducting
  system Eq.\ \eqref{h2} placed on a long strip with open boundaries in the
  $y$-direction (width 52 sites) and periodic boundary conditions along $x$. The
  spectral function is averaged over the lower half of the strip (sites
  1-26) to clearly show the spectrum associated with one edge. Note
  that there are two distinct edge modes in the chiral
  $p_+^\uparrow\otimes p_+^\downarrow$ phase shown in panel (c), consistent
  with $C=2$. They become more clearly visible in panel (d) where the
  degeneracy is lifted by SOC.  We use  $\eta=0.2$,
  $\mu=-2.1$, $\Delta_\sigma=0.5$ and $\lambda_R$ as indicated. }
  \label{fig3}
\end{figure}

\subsection{Paired state}

It is clear that for
weak attractive interactions the type of FS depicted in Fig.\
\ref{fig2}(a)  cannot support the
conventional spin-singlet SC state with zero-momentum Cooper
pairs. Except for the 4 points where the two Fermi surfaces intersect,
to form a zero-momentum Cooper pair 
an electron with crystal momentum $+\bk$ must pair with the {\em same
  spin} electron at momentum $-\bk$. Such an equal-spin pairing requires an
orbital pair wave-function that is odd under inversion. On the square lattice the
simplest possibility is the $p$-wave state. Furthermore, to maximize the
condensation energy the paired state should be fully gapped which
leads directly to the chiral $p_\pm$ pair state in each spin channel
as the most natural SC ground state compatible with the altermagnetic
normal metal Eq.\ \eqref{h0}.
This conclusion is consistent with the results of Ref.\
\cite{Zhu2023} where a similar model was solved self-consistently and
with our own computations \cite{Sunny2024}.

The SC state can be described using a four-component  Nambu spinor
$\Psi_\bk=(c_{\bk\uparrow}, c_{\bk\downarrow}, 
c_{-\bk\uparrow}^\dagger,- c_{-\bk\downarrow}^\dagger)^T$ and  the Bogoliubov-de Gennes (BdG)
Hamiltonian of the form
\begin{equation}\label{h2}
  \cH=\sum_\bk\Psi_\bk^\dagger
  \begin{pmatrix}
    h_\bk & \hat\Delta_\bk \\
    \hat\Delta_\bk^\dagger & -h_\bk^\ast
    \end{pmatrix}
  \Psi_\bk,
\end{equation}
with $h_\bk=h_0(\bk)+h_R(\bk)-\mu$ and the pair field matrix
$\hat\Delta_\bk={\rm diag}(\Delta_{\bk\uparrow}, \Delta_{\bk\downarrow})$. Here 
\begin{equation}\label{h3}
 \Delta_{\bk\sigma}=\Delta_{\sigma}(\sin{k_x}\pm i\sin{k_y}) 
\end{equation}
are chiral $p$-wave order parameters for spin $\sigma$ electrons and
$\mu$ denotes the chemical potential.

In
the absence of SOC the two spin channels are uncoupled and the
$4\times 4$ BdG matrix in Eq.\ \eqref{h2} becomes block-diagonal in
spin space. Each block then describes an independent  $p_x\pm ip_y$  superconductor
for a given spin projection, with a fully gapped excitation spectrum 
\begin{equation}\label{h4}
 E_{\bk\sigma}=\sqrt{(\epsilon_{\bk\sigma}-\mu)^2+|\Delta_{\bk\sigma}|^2},
\end{equation}
where $\epsilon_{\bk\sigma}$ denote the normal-state excitation
energies of spin-$\sigma$ electrons that follow from Eq.\ \eqref{h0}. 
Viewed in isolation each such
superconductor is topological, characterized by a nonzero BdG Chern
number $C=\pm 1$ for $p_x\pm ip_y$, respectively. Of the 4 degenerate
ground states of the combined spin up/down sectors two are chiral,
namely $p_+^\uparrow\otimes p_+^\downarrow$ and $p_-^\uparrow\otimes
p_-^\downarrow$, with the total Chern number $C=+2$ and $-2$,
respectively.  The remaining two are helical and have
$C=0$. Correspondingly, when placed on a long strip their excitation
spectra exhibit chiral or helical gapless edge modes shown in Fig.\
\ref{fig3}(a,c). Note that the 4-fold degeneracy mentioned above
simply refers to 4 possible combinations of $+/-$ signs in Eq.\
\eqref{h3}  and is unrelated to the topological degeneracy on a torus
that we discuss below.  

The presence of SOC ($\lambda_R\neq 0$) alters the above picture in two
important ways. First, in the absence of inversion symmetry one can no
longer classify pairing as even or odd under inversion. Instead, the
two channels 
can mix. For weak Rashba SOC we thus expect the pair field matrix
$\hat\Delta$ to acquire small off-diagonal elements that represent
spin-singlet components of the SC order parameter allowed
by symmetry. Second, the Rashba term couples spin up and down sectors of
the BdG Hamiltonian Eq.\ \eqref{h2}. This has the effect of splitting
the four-fold ground state degeneracy mentioned above. We find,
consistent with the results of Ref.\ \cite{Zhu2023}, that Rashba SOC favors
one of the helical states with total Chern number $C=0$, specifically
$p_-^\uparrow\otimes p_+^\downarrow$. Intuitively, this
can be understood as follows. Microscopically, the Rashba term
originates from the atomic SOC term $\bS\cdot\bL$ which conserves the
total angular momentum $\bJ=\bS+\bL$. It is easy to see that
$p_-^\uparrow\otimes p_+^\downarrow$ 
is compatible with this conservation law while the other three states 
states are not. Specifically, $J_z$ of both components in
$p_-^\uparrow\otimes p_+^\downarrow$  vanishes (for instance a
$p_-^\uparrow$ Cooper pair has $s_z=2\times (1/2)$ and $l_z=-1$) and
hence Rashba induced transfer of a Cooper pair from  $p_-^\uparrow$ to
$p_+^\downarrow$ conserves the total angular momentum $\bJ$.

According to the above discussion in the presence of a small 
but nonzero $\lambda_R$, our system has a unique ground
state characterized as $p_-^\uparrow\otimes p_+^\downarrow$. A
detailed calculation supporting this conclusion is presented in
Appendix A, where we also discuss the effect of proximity induced spin
singlet pairing and applied in-plane magnetic field $B_x$. The former
also gaps out the helical 
edge modes but, in contrast to SOC, admits two degenerate ground
states, namely $p_-^\uparrow\otimes p_+^\downarrow$ and
$p_+^\uparrow\otimes p_-^\downarrow$. Zeeman field, on the other hand, is
found to select the two chiral phases as ground states. We thus
conclude that by applying perturbations listed above one can
efficiently  control the system and select any of the four degenerate
ground states, making the SC altermagnet a flexible and potentially useful platform
for explorations of topological superconductivity. 

We finally note an important case of spontaneously generated
SC gap at the edge of the helical phase
in the absence of any perturbation ($\lambda_R=B_x=0$) or proximity
effect. Because the edge breaks the inversion symmetry one expects
pairing near the edge to mix $p$ and $s$ channels and hence permit
opposite-spin pairing, even if the bulk is purely equal-spin $p$-wave.      
Indeed the
two helical edge modes displayed in Fig.\ \ref{fig3}(c) belong to the 
opposite spin projections $\uparrow$ and $\downarrow$. In the presence
of attractive interactions (e.g.\ those  responsible for the bulk $p$-wave
SC order) these low-energy modes will be susceptible to the formation
of Cooper pairs composed of the time-reversed edge states $(+k,\uparrow)$
and $(-k,\downarrow)$. The 
corresponding term in the BdG Hamiltonian will be
\begin{equation}\label{h4a}
\delta\cH=\sum_{k,y}\left[\Delta^{\rm
  edge}_k(y)c_{k\uparrow}^\dagger(y) c_{-k\downarrow}^\dagger(y)+{\rm h.c.}\right]
\end{equation}
with $k$ denoting the momentum along the edge and $y$ the
perpendicular coordinate.  As we will demonstrate below, non-zero $\Delta^{\rm
  edge}_k(y)$ opens a gap in the helical modes. Importantly, it also
facilitates coupling between spin-up and spin-down sectors of the
theory, hence enabling the torus/annulus correspondence indicated in
Fig.\ \ref{fig1}.

Edge modes in the two chiral phases are protected by the bulk
topological invariant and hence cannot be gapped by any
perturbation that does not close the bulk gap.
For this reason we expect the two helical phases with
gapped edges (due to $\Delta^{\rm edge}$)  to be energetically favored
over the chiral phases even in the absence of any other symmetry
breaking perturbations, such as SOC.

\subsection{Symmetries}

The BdG Hamiltonian in Eq.\ \eqref{h2} respects the charge conjugation
symmetry $\cC$ generated by $\sigma_z\tau_x
H_{-\bk}^\ast\tau_x\sigma_z=-H_\bk$ with $\cC^2=+1$. Here $H_\bk$
refers to the $4\times 4$ BdG matrix in Eq.\ \eqref{h2} and $\tau$ are
Pauli matrices in the Nambu space. This places the system in the
Altland-Zirnbauer class D which has an integer topological
classification in two dimensions; the relevant index is just the BdG
Chern number mentioned above which is indeed non-zero for the two
chiral ground states.

In the absence of SOC the two helical states
respect an additional antiunitary symmetry $\cT'=C_4\cT$ with
$\cT'^2=-1$. The system is then in class DIII which has $\mathbb{Z}_2$ topological
classification in two dimensions. However, because the edge breaks the
$C_4$ rotation symmetry the corresponding $\mathbb{Z}_2$ index alone
cannot protect the Kramers-like degeneracy between the two edge modes
at $k=0$, clearly visible in Fig.\ \ref{fig3}(c). The edge mode crossing
is instead protected by the $\mathbb{Z}_2^\uparrow\times
\mathbb{Z}_2^\downarrow$ number parity conservation symmetry that is
a remnant of the  $U(1)^\uparrow\times U(1)^\downarrow$  electron
number conservation present in the normal state. 

Simply put, as long as the two spin blocks remain decoupled the
crossing of the two helical edge modes will be protected. Any
perturbation that couples the two spin sectors breaks the $\mathbb{Z}_2^\uparrow\times
\mathbb{Z}_2^\downarrow$ symmetry down to a single $\mathbb{Z}_2$
associated with the global number parity conservation and will
generically open a gap.

Indeed we see in Fig.\ \ref{fig3}(d) that  $\lambda_R\neq 0$ opens up a
gap in the edge modes at
$k=0$. Adding a small spin-singlet component to the SC pair matrix
likewise couples the two spin projections, breaks the $\mathbb{Z}_2^\uparrow\times
\mathbb{Z}_2^\downarrow$ symmetry by the same mechanism and gaps out
the helical edge modes. Applied in-plane magnetic field that acts as a
Zeeman term $\bB_\parallel\cdot\bsig$ also breaks the symmetry 
and opens a gap if $\bB_\parallel$ has a non-zero projection onto the
edge (otherwise a mirror symmetry protects the helical crossing) .

\section{Topological degeneracy on torus}

In this Section we summarize the known facts about topological
degeneracy of superconductors. We review both conventional
spin-singlet $s$-wave and equal-spin chiral $p$-wave cases. According
to the modern theory \cite{Hansson2004} both are characterized by topological order when
placed on a torus with a subtle difference distinguishing $s$- and $p$-wave
paired states, closely related to the non-Abelian exchange statistics
of vortices in the latter \cite{Oshikawa2006}.  This is one of the key properties that
chiral $p$-wave SC shares with the Moore-Read Pfaffian fractional
quantum Hall state \cite{Read2000,Oshikawa2007}. 

\subsection{General considerations}

According to the modern understanding superconductors are
topologically ordered phases of matter \cite{Hansson2004}.
The ground-state degeneracy is
associated with the even/odd number of superconducting magnetic flux
quanta $\Phi_0=hc/2e$ threaded through the two holes of the torus. To
illustrate this in the simplest possible setting we consider
weak-coupling continuum 
BCS theory defined by the BdG Hamiltonian of the form
\begin{equation}\label{h5}
  \cH=\sum_\bk\Psi_\bk^\dagger
  \begin{pmatrix}
    \xi_\bk & \Delta_\bk \\
    \Delta_\bk^\ast & -\xi_\bk^\ast
    \end{pmatrix}\Psi_\bk,
  \end{equation}
where $\xi_\bk=\hbar^2k^2/2m-\mu$ is the kinetic energy referenced to
  the chemical potential $\mu$. For spin-singlet $s$-wave we have
\begin{equation}\label{h6}
  \Delta_\bk=\Delta_s, \ \ \ \ 
\Psi_\bk=
  \begin{pmatrix}
    c_{\bk\uparrow} \\ c_{-\bk\downarrow}^\dagger
\end{pmatrix},
\end{equation}
while for equal-spin chiral $p$-wave
\begin{equation}\label{h7}
  \Delta_\bk=\Delta_p(k_x-ik_y), \ \ \ \ 
 \Psi_\bk=
  \begin{pmatrix}
    c_{\bk} \\ c_{-\bk}^\dagger
 \end{pmatrix}.
\end{equation}

  The toroidal geometry is implemented by imposing periodic boundary conditions
  along both $x$ and $y$ directions
  which, for a system of size $L_x\times L_y$, implies discrete values
  for the allowed   momenta
\begin{equation}\label{h8}
 \bk=2\pi(n_x/L_x,n_y/L_y), \ \ \ \  n_x,n_y\in \mathbb{Z}. 
\end{equation}
Magnetic fluxes $(\Phi_x,\Phi_y)$ threaded through the torus holes are
included by the minimal substitution in the kinetic  term,   $\bk\to
-i\nabla -(e/\hbar c)\bA$  where $\bA$ is the vector potential satisfying 
\begin{equation}\label{h9}
\oint_{\Omega_j}\bA\cdot d{\bm l}=\Phi_j,
  \end{equation}
 with ${\Omega_j}$ denoting a non-contractible loop on the torus. We
 will work in the gauge $\bA=(\Phi_x/L_x,\Phi_y/L_y)$ which preserves
 the translational invariance of the original problem.

 When the magnetic flux is non-zero the superconductor will carry
 persistent  current with density
\begin{equation}\label{h99}
 \bj_s=\rho_s\left( \nabla\varphi-{2e\over \hbar c}\bA\right),
  \end{equation}
where $\rho_s$ denotes the superfluid density.  Such a current  incurs an 
 energy cost $\propto\bj_s^2$ per unit area. In Eq.\ \eqref{h99} $\varphi$
 denotes the phase of the 
 pair field $\Delta_s(\br)=\Delta_{0}e^{i\varphi(\br)}$ which is allowed to vary along
 the torus so as to minimize the system energy. Importantly, such spatial
 variations are  constrained by 
 single-valuedness of $\Delta_s(\br)$ which implies that
 $\varphi(\br)$ can wind by an integer multiple of $2\pi$ along each
 non-contractible loop. For $p$-wave it is necessary to symmetrize the
 placement of the phase, 
$\Delta_p(\br)=\Delta_0 e^{i\varphi(\br)/2}(-i\partial_x+\partial_y)
e^{i\varphi(\br)/2}$, in order to ensure full gauge invariance of the
theory \cite{Vafek2001,Liu2015}.
\begin{figure*}[t]
  \includegraphics[width=16.5cm]{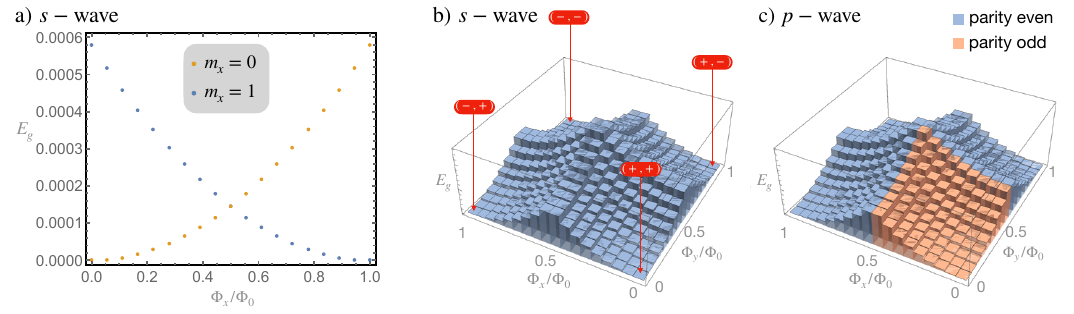}
  \caption{Ground state energy $E_g$ of (a-b) $s$-wave and (c)
    weak-pairing chiral
    $p$-wave superconductor on a torus as a function of magnetic fluxes
  $\Phi_x$ and $\Phi_y$ threaded through its holes. We use the lattice
model with $t=1$, $\mu=-2.1$ and $\Delta_0=0.1$. The results are
insensitive to exact values, except for parity in (b) which requires
$-4<\mu <+4$ to remain in the weak-pairing phase. }
  \label{fig4}
\end{figure*}

A non-zero phase winding can be thought of as
 inserting a vortex into one of the torus holes. Mathematically, we
 can describe it as
\begin{equation}\label{h10}
  \varphi(\br)=2\pi\br\cdot(m_x/L_x,m_y/L_y)
  \end{equation}
where integers $(m_x,m_y)$ count vortices in two holes.
 A vortex will enter when it lowers the system ground state energy. If
we slowly ramp up the  flux $\Phi_x$ from zero the 
increasing supercurrent will drive the  energy up in proportion to 
$\bj_s^2$. When $\Phi_x$ exceeds $\Phi_0/2$ Fig.\ \ref{fig4}(a) shows
that it becomes
energetically favorable to insert a vortex and partially screen the
supercurrent. At $\Phi_x=\Phi_0$ this screening becomes complete and
no current flows once again by virtue of the prefect cancellation of the vector
potential term in Eq.\ \eqref{h99} by $\nabla\varphi$.

It is useful to bring the BdG  Hamiltonian in the presence of vortices
back into translation-invariant form. This is achieved by a ``large'' gauge transformation on the electron
 operators
\begin{equation}\label{h11}
 \Psi_\br=
  \begin{pmatrix}
    c_{\br\uparrow} \\ c_{\br\downarrow}^\dagger
  \end{pmatrix} \to
    \begin{pmatrix}
    e^{i\varphi(\br)}  c_{\br\uparrow} \\ c_{\br\downarrow}^\dagger
        \end{pmatrix} .
  \end{equation}
and similar for $p$-wave. In this representation the Hamiltonian Eq.\
\eqref{h5} can be written in momentum space as
\begin{equation}\label{h12}
 H_\bk=
  \begin{pmatrix}
    \xi_{\bk-\ba+2\bv} & \Delta_{\bk+\bv} \\
    \Delta_{\bk+\bv} & -\xi_{\bk+\ba}
    \end{pmatrix},
  \end{equation}
where we defined
\begin{align}\label{h13}
  \ba&=(e/\hbar c)\bA = \pi(\Phi_x/L_x,\Phi_y/L_y)/\Phi_0, \nonumber \\
  \bv&={1\over 2}\nabla\varphi=\pi(m_x/L_x,m_y/L_y), 
\end{align}
and both $\xi_\bk$ and $\Delta_\bk$ are now real. Note that Eq.\ \eqref{h12}
is valid for both $s$- and $p$-wave order parameters

 In the special case when the magnetic flux is an integer multiple of
 $\Phi_0$ and matches the number of vortices, that is when
\begin{equation}\label{h14}
  (\Phi_x,\Phi_y)=\Phi_0(m_x,m_y),
\end{equation}
we have $\ba=\bv=\pi(m_x/L_x,m_y/L_y)$ and the
Hamiltonian Eq.\ \eqref{h12} becomes 
\begin{equation}\label{h15}
 H_\bk=
  \begin{pmatrix}
    \xi_{\bk+\bv} & \Delta_{\bk+\bv} \\
    \Delta_{\bk+\bv} & -\xi_{\bk+\bv}
    \end{pmatrix}.
\end{equation}
We observe that insertion of magnetic fluxes that satisfy Eq.\
\eqref{h14}, leads to a simple shift of momenta $\bk\to\bk+\bv$ in
the original zero-flux Hamiltonian Eq.\ \eqref{h5}.  For a large system this
explains why all such states are degenerate in energy. Fundamentally, 
this is nothing  but the well-known statement of flux-quantization:
superconductor is able to screen the external magnetic field threaded
through a hole by adjusting its phase as long the flux is an integer
multiple of the SC flux quantum $\Phi_0$.

There is, however, an important distinction between even and odd
number of SC flux quanta threading the torus which lies at the root of the topological
degeneracy. For even $m_j$ the new allowed values of momenta are
simply shifted by one unit and continue to satisfy Eq.\
\eqref{h8} with $n_j$ integral. Hence, the electron wavefunctions are identical
to those in the absence of flux. For odd $m_j$, however, the momenta are
shifted by {\em half a unit}, meaning that the Hamiltonian is now
equivalent to Eq.\ \eqref{h5}  with one or both of $n_j$ in Eq.\ \eqref{h8} half-integral. In real space this corresponds to antiperiodic
boundary conditions for the wavefunctions which are therefore
distinct from the zero flux case.   
This gives rise to four ground states that we label by a pair of indices
\begin{equation}\label{h15b}
(\nu_x,\nu_y)=((-1)^{m_x}, (-1)^{m_y}),
\end{equation}
with distinct and mutually orthogonal many-body wavefunctions.

As already mentioned, in the
thermodynamic limit the four ground states become
asymptotically degenerate because the total energy in a large system
is not expected to depend on periodic versus antiperiodic boundary
conditions. The degeneracy is topological in the sense that no local
measurement made within the 2D system can distinguish between
the ground states; to detect the fluxes one must perform non-local measurements
along two non-contractible loops of the torus.

\subsection{$s$-wave state}

To illustrate the above general considerations we calculate the ground
state energy of an $s$-wave superconductor placed on a torus in the
presence of arbitrary fluxes $(\Phi_x,\Phi_y)$. This is obtained by
summing all negative eigenvalues of the BdG Hamiltonian Eq.\
\eqref{h12} and minimizing the result over the number of inserted
vortices. For the numerical evaluation it is easier to revert to the lattice
model with $\xi_\bk=-2t(\cos{k_x}+\cos{k_y})-\mu$.  
The result of this computation is given in Fig.\
\ref{fig4}(a) and clearly shows minima at integer fluxes
corresponding to four degenerate ground states labeled in accord with
Eq.\ \eqref{h15b} as $(+,+)$,
$(+,-)$, $(-,+)$ and $(-,-)$.    

\subsection{Chiral $p$-wave state}

The analysis for chiral $p$-wave proceeds along the same
path and leads to the same conclusion of
four-fold ground state degeneracy on a torus.
There is, however, one important difference: Whereas in the $s$-wave SC
all four degenerate ground states belong to the even parity subspace,
for chiral $p$-wave in the weak pairing phase one of the ground
states, namely $(+,+)$, belongs to the odd-parity subspace \cite{Read2000}. As we
shall see this subtle distinction is fundamentally related to the
non-Abelian exchange statistics of vortices in the chiral $p$-wave phase. 

To understand the significance of the above statement recall that
while BdG Hamiltonians do not
conserve electron number they do conserve the number parity. Hence any
eigenstate can be classified as parity even or parity odd. The BCS
many-body ground state (for the $p$-wave case) can be written as  
\begin{equation}\label{h16}
|\Psi_G\rangle={\prod_\bk}'\left(u_\bk+v_\bk c^\dagger_\bk c^\dagger_{-\bk}\right)|0\rangle,
\end{equation}
where the prime indicates that each pair is to be included once.
$(u_\bk,v_\bk)$ denote the BCS coherence factors which satisfy
\begin{equation}\label{h17}
  |u_\bk|^2={1\over 2}\left(1+{\xi_\bk\over E_\bk}\right), \ \ 
  |v_\bk|^2={1\over 2}\left(1-{\xi_\bk\over E_\bk}\right), 
\end{equation}
with $E_\bk=\sqrt{\xi_\bk^2+|\Delta_\bk|^2}$ and
$u_\bk/v_\bk=-(E_\bk-\xi_\bk)/\Delta_\bk^\ast$.

In the $(+,+)$ sector the product in Eq.\ \eqref{h16} includes the $\bk=0$
term. (The other three sectors obey antiperiodic boundary condition
in at least one direction which implies that the $\bk=0$ term is
absent from the product.)
Importantly, $\Delta_\bk$ vanishes at $\bk=0$ which according to Eq.\
\eqref{h17}  implies that the corresponding coherence factor 
$v_{\bm 0}$ also vanishes. This is a manifestation of the simple fact that under
equal-spin pairing there is only a single electron state available at
$\bk=0$ and one therefore cannot form a Cooper pair from electrons at $+\bk$ and $-\bk$.
Instead, one must  consider the occupancy of this state
separately. The single state at $\bk=0$ will be
occupied provided $\mu>0$ which defines the weak-pairing phase of the
model \cite{Read2000}. The ground-state wave-function in the weak-pairing phase hence reads
\begin{equation}\label{h18}
|\Psi_G^{(+,+)}\rangle={\prod_\bk}'\left(u_\bk+v_\bk c^\dagger_\bk
  c^\dagger_{-\bk}\right) c^\dagger_{\bm 0}|0\rangle,
\end{equation}
and belongs to the odd parity subspace of the theory. The other three
ground states, namely $(-,+)$, $(+,-)$ and $(-,-)$, are described by
Eq.\ \eqref{h16} and belong to the even parity subspace. In the
following we will refer to this feature as the ``3:1 parity rule''.

In the strong
pairing phase ($\mu<0$) all four sectors are described by
Eq.\ \eqref{h16} and belong to the even parity subspace. Ground
states of a spin-singlet $s$-wave SC discussed in the previous
subsection likewise belong to the even parity subspace.

Fig.\ \ref{fig4}(b) shows the ground state energy, calculated for the
lattice model with the pair field given by Eq.\ \eqref{h3}. As for
$s$-wave we observe four degenerate ground states at integer
fluxes. However, this being a weak-pairing chiral $p$-wave phase, we
now have an odd-parity ground state in the quadrant containing the
$(+,+)$ absolute minimum. 

The discussion given above may give an impression that the 3:1 parity
rule of the ground state manifold in the weak pairing phase of a $p_x+ip_y$
superconductor is a fragile property because it relies e.g.\ on the perfect translation
invariance that allows a description in terms of the crystal momentum. However,
this is not the case. The 3:1 rule  is a robust
topological  property of the weak pairing phase and persists in the
presence of disorder and other perturbations that do not close
the gap between the ground state manifold and the excited states. This
can be established by an adiabatic continuity argument and also,
perhaps more instructively, by appealing to the Oshikawa-Senthil
construction \cite{Oshikawa2006, Oshikawa2007}
which, based on earlier works \cite{Verlinde1988,Einarsson1990},
establishes a fundamental relation between the ground state
degeneracy on a torus and the existence of fractionalized
excitations.  Specifically, this powerful argument shows that one can
cycle between the individual ground 
states by creating a pair of such fractional excitations, dragging one of them
along a non-contractible loop and then again annihilating the pair.

Applied to the problem at hand the 3:1 property can be understood
as a fundamental 
consequence of the non-Abelian nature of vortex/antivortex excitations
in the weak pairing phase of a chiral $p$-wave superconductor and
mirrors that of the Moore-Read 
fractional quantum Hall phase. It turns out that,  if one attempts to
create the $(+,+)$ state starting from any
of the even-parity ground states, the above procedure instead leads to an
excited state with an extra Bogoliubov quasiparticle above the $(+,+)$
ground
state. Removing this quasiparticle then gives the odd-parity ground
state adiabatically connected to Eq.\ \eqref{h18}. This conclusion is
general and does not require 
translation-invariant system. In the next Section we shall demonstrate 
that both the ground-state degeneracy and the 3:1 parity effect characteristic of the
weak-pairing phase  occur in the model
introduced in Sec.\ II when placed on an annulus and persist in the
presence of disorder.
\setlength{\tabcolsep}{6pt} 
\begin{table*}[t]
\centering 
\caption{Parity $P$ and ground state energy splitting parameter 
  $\varepsilon^{(\nu_x,\nu_y)}$ for the
  four ground state sectors in the long strip geometry. $N_x=64$ in
  all cases while $N_y$ varies as indicated. We use  $\eta=0.2$,
  $\mu=-2.1$, $\Delta_{p\sigma}=0.5$, $\Delta_{sj}=\pm 0.2$ and $\lambda_R=0$.}
\begin{tabular}{c c c c c}
\hline\hline
  $(\nu_x,\nu_y)$ &\  ~\ $(+,+)$\  ~\  ~\ &\  ~\ ~\ $(+,-)$ \  ~\ ~\   &\  ~\ ~\ $(-,+)$ ~\ ~\ &\ 
                                                    ~\ ~\ $(-,-)$ ~\ ~\ \\
\hline
$P$ &            $+$      &  $-$ &  $+$  &  $+$ \\
\hline
$\varepsilon\ (N_y=16)$  & $-5.15163\times 10^{-10}$ & $-3.93501\times 10^{-10}$  & $3.93491\times 10^{-10}$
                                                             &  $5.15174\times 10^{-10}$ \\
\hline
$\varepsilon\ (N_y=32)$  & $-2.26019\times 10^{-10}$ & $-2.26019\times 10^{-10}$  & $2.26019\times 10^{-10}$
                                                             &
                                                               $2.26019\times 10^{-10}$
\\
  \hline
$\varepsilon\ (N_y=64)$  & $-1.12745\times 10^{-10}$ & $-1.12745\times 10^{-10}$  & $1.12745\times 10^{-10}$
                                                             &
                                                               $1.12745\times 10^{-10}$ \\
\hline\hline
\end{tabular}
\label{table1}
\end{table*}

\section{Topological degeneracy on annulus}

We now consider the $p_-^\uparrow\otimes
p_+^\downarrow$ phase of the model introduced in
Sec.\ II and show that, when placed on an annulus with the appropriate
symmetry breaking edge perturbations, it exhibits the ground-state
degeneracy characteristic of the chiral $p$-wave superconductor on a
torus. For reasons that will become clear momentarily we consider a
very weak uniform Rashba SOC to stabilize the helical $p_-^\uparrow\otimes
p_+^\downarrow$ phase, and include a spontaneously generated $s$-wave order
parameter near the edges to play the role of  the symmetry breaking
edge perturbation.
\begin{figure}[t]
  \includegraphics[width=8.5cm]{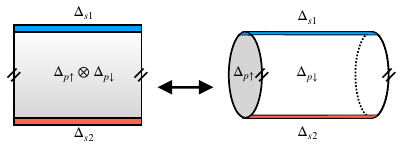}
  \caption{$p_-^\uparrow\otimes p_+^\downarrow$ state on a strip with
    spin-singlet SC order spontaneously nucleated near its horizontal
    edges (left) is electronically equivalent to chiral $p$-wave on a
    cylinder (right). With periodic boundary conditions along $x$ this
  has a topology of a torus.}
  \label{fig5}
\end{figure}

\subsection{Long strip geometry}
As a warm-up exercise we first place the $p_-^\uparrow\otimes
p_+^\downarrow$ superconductor on a long strip with periodic boundary
conditions (b.c.) along $x$ and open along $y$, illustrated in Fig.\
\ref{fig5}.  According to our discussion
in Sec.\ I, when equipped with symmetry breaking perturbations along
the open edges, this configuration can be thought of as having a
topology of the torus. The two spin sectors, decoupled in the bulk of
the strip, furnish the opposite  walls of the torus while the symmetry breaking
perturbations along the edges serve to glue the two walls together by
allowing electrons to flip their spin and hence circle around the full
cycle. While this configuration may not be easy to realize in the
laboratory (due to periodic boundary conditions along $x$) it is
instructive to consider theoretically because of its translation
invariance in one direction. In the next subsection we will consider a
true annulus which can 
be easily fabricated and shows identical behavior.

To model this geometry we adapt Hamiltonian Eq.\ \eqref{h2}
by partially Fourier transforming to real space along $y$ and add the
edge term Eq.\ \eqref{h4a} to facilitate transfer of electrons between
the spin sectors. For simplicity we assume $\Delta^{\rm edge}_k$ to be real,
$k$-independent and equal to $\Delta_{s1,2}$ as indicated in  Fig.\
\ref{fig5}. This choice can be understood by considering the
corresponding Ginzburg-Landau free energy, which in the absence of SOC
contains, to the lowest order, the following terms coupling edge order
parameters to the bulk,
\begin{equation}\label{h19}
  -\beta_{sp}\sum_{j=1,2}\left(\Delta_{sj}^2\Delta_{p\uparrow}^\ast\Delta_{p\downarrow}^\ast
  + {\rm c.c.}\right),
\end{equation}
with $\beta_{sp}$ a positive constant. Bilinear terms, such as
$\Delta_{sj}\Delta_{p\uparrow}^\ast$, are not allowed because they
are odd under $x\to -x$ in the geometry under consideration. Note that
adding weak SOC breaks this symmetry and we will discuss this effect
separately.

Eq.\ \eqref{h19} reveals an important sign ambiguity: switching
$\Delta_{sj}\to -\Delta_{sj}$ has no effect on the GL free energy. In
addition, Eq.\ \eqref{h19} implies that we can always choose $\Delta_{sj}$ to be
real if we also choose $\Delta_{p\uparrow}\Delta_{p\downarrow}$ as real. This follows from
inspecting the corresponding Josephson energy, proportional to
$-\cos(2\varphi_{sj}-\varphi_{p\uparrow}-\varphi_{p\downarrow}) $
where $\varphi$ denote various order parameter phases. The Josephson
energy is minimized by taking
$\varphi_{p\uparrow}=-\varphi_{p\downarrow}$ and $\varphi_{sj}=0,\pi$ 
We will see below that while individual signs of $\Delta_{sj}$ are
arbitrary, the sum of their phases $\varphi_{s1}+\varphi_{s2}$ has physical
significance; it plays the role of  the Aharonov-Bohm-like phase of an electron
traversing the $y$-cycle of the torus. 

Periodic vs.\ antiperiodic boundary conditions along
$x$ are easily implemented by choosing integer or half-integer
crystal momenta $k$. Furthermore, a brief reflection reveals that
taking $\Delta_{s1}\Delta_{s2}>0 \ (<0)$ corresponds to periodic
(antiperiodic) boundary conditions along $y$. This is because to
complete a full cycle along $y$ an electron must traverse each edge
once. When $\Delta_{s1}\Delta_{s2}>0$ the total phase acquired is $0$
or $2\pi$ while for $\Delta_{s1}\Delta_{s2}<0$ it is $\pm\pi$, the
latter corresponding to antiperiodic b.c.

To demonstrate the electronic equivalence between the strip and the
torus we numerically evaluate the ground state energy $E_g$ of the
strip in the four flux sectors as well as the corresponding electron
number parity $P$. According to Ref.\ \cite{Halperin2015} the later can be calculated
from the knowledge of BdG eigenvectors $\phi_{nk}=(u_{nk},v_{nk})^T$ belonging to
positive energy eigenvalues $E_{nk}$. Specifically, for the strip
geometry at hand the parity is given by $P=\prod_k P_k$ with       
\begin{equation}\label{h20}
P_k=(-1)^{{\rm rank}(V_k)}
\end{equation}
where $V_k$ is a $(2N_y\times 2N_y)$ matrix of column vectors $v_{nk}$ and
$N_y$ denotes the width of the strip.

Our results are summarized in Table I. We find four nearly-degenerate
ground states corresponding to four sectors labeled by
$(\nu_x,\nu_y)$. Three of these belong to the even parity subspace and
one to the odd parity subspace. In contrast to the chiral $p$-wave on
a torus we find that $(+,-)$ is the odd-parity ground state. This can
be understood as follows: An electron traversing the $y$-cycle of the
strip must flip its spin twice. This can be thought of as a $2\pi$
rotation in the spin space which entails additional Berry phase
$\pi$. Hence, effectively, the $(+,-)$ sector provides periodic
boundary conditions along both cycles and hosts the lone odd-parity
ground state.
\begin{figure*}[t]
  \includegraphics[width=17.5cm]{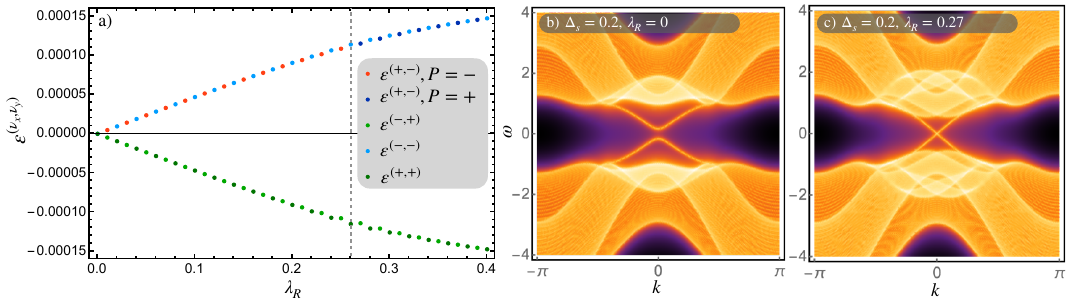}
  \caption{Splitting of the ground-state degeneracy in the presence of
    Rashba SOC. Panel (a) shows the fractional energy splitting
    $\varepsilon^{(\nu_x,\nu_y)}$ as a function of $\lambda_R$. The
    vertical dashed line indicates the transition between even and odd parity of the
    $(+,-)$ state. Panels (b,c) display edge spectral function $A(\omega,k)$ for
    $\lambda_R=0.0$ and $0.27$, the latter marking the closing of the
    edge energy gap. The same parameters are used as in Table I with $(N_x,N_y)=(64,32)$. }
  \label{fig6}
\end{figure*}

We quantify the ground state degeneracy by parameter
$\varepsilon^{(\nu_x,\nu_y)}$ defined as the relative fractional
energy difference between $E_g^{(\nu_x,\nu_y)}$ and the average of the
four ground state energies 
$E_g^{\rm avg}$,
\begin{equation}\label{h20b}
\varepsilon^{(\nu_x,\nu_y)}=\left(E_g^{(\nu_x,\nu_y)}-E_g^{\rm avg}\right)/|E_g^{\rm
  avg}|.
\end{equation}
For $N_x=64$ and any $N_y\geq 32$ the two pairs of
  ground states with the same $\nu_x$ are found to be exactly
  degenerate within our numerical accuracy $\sim 10^{-12}$. The
  fractional energy splitting between these two pairs is at a
  $\sim 10^{-10}$ level and is furthermore seen to be decreasing in proportion to
  $1/N_y$. We thus conclude that the four ground states are indeed
  exactly degenerate in the thermodynamic limit and obey the 3:1 parity
  rule characteristic of the chiral $p$-wave superconductor on a torus.   

In the presence of Rashba SOC the bulk  $\mathbb{Z}_2^\uparrow\times
\mathbb{Z}_2^\downarrow$ symmetry breaks down to a single
$\mathbb{Z}_2$ global parity conservation. Physically, SOC enables Cooper pairs to `tunnel' between spin-up and spin-down sectors of the
theory away from the edges of the strip.
One thus expects non-zero $\lambda_R$ to weaken the strip-torus
correspondence indicated  in Fig.\ \ref{fig5} and lift  the exact degeneracy
between the four ground states.  Fig.\ \ref{fig6}
quantifies this splitting. We observe that  increasing $\lambda_R$ indeed
splits the ground-state manifold into two nearly degenerate pairs of
states labeled by common $\nu_y$. The split between  $\nu_y=+$ and
$-$ states grows approximately linearly with $\lambda_R$ but remains at least
two orders of magnitude below all other energy scales (e.g.\ the bulk
excitation gap). For $\lambda_R < \lambda_{Rc}\simeq 0.27$ the system also
continues to obey the 3:1 parity rule; for larger $\lambda_R$ all
four ground states belong to the even parity subspace. As shown in panels (b,c)
this transition is accompanied by closing of the edge gap. When the gap
reopens for larger $\lambda_R$ the system on the strip can no longer be thought of
as topologically equivalent to the torus.

The above behavior can be understood as follows. We know that  either $\Delta_s$ or
Rashba SOC open a gap in the helical edge modes. When both are present
the two gaps compete. For $\lambda_R\lesssim \Delta_s$ the edge gap is
dominated by the pairing term which we rely on to implement periodic vs.\
antiperiodic boundary conditions in the $y$-direction. In this regime the system
is sensitive to the sign of $\Delta_{s1}\Delta_{s2}$ and  behaves,
effectively, as a chiral $p$-wave SC on a torus. In the opposite regime
the edge gap is dominated by SOC which imposes periodic b.c.\ along $y$,
independent of $\sgn(\Delta_{s1}\Delta_{s2})$. We checked that
modifying the Rashba SOC such that $\lambda_R$ changes sign between
the two edges restores the 3:1 parity behavior in this
regime. However, it is not clear how this could be implemented
experimentally.

We conclude that the ground state degeneracy
characteristic of chiral $p$-wave SC on a torus can indeed be realized in the
helical  $p_-^\uparrow\otimes p_+^\downarrow$  phase of the model
placed on the long strip, provided that Rashba SOC is small compared
to the SC gap at the edges.
\begin{figure*}[t]
  \includegraphics[width=17.5cm]{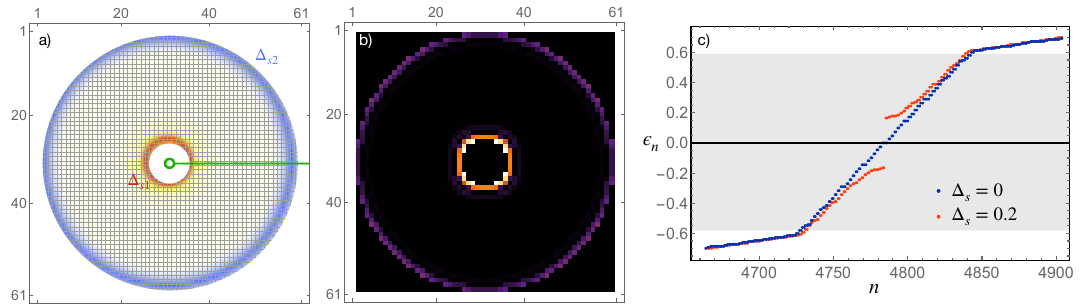}
  \caption{The $p_-^\uparrow\otimes p_+^\downarrow$ state on an
    annulus. a) Square lattice cluster containing 2392
    sites used to model an annulus with   inner radius $R_1=5$ and outer radius $R=28$. Bonds crossing the vertical green line have their hopping
    and pairing amplitudes reversed when implementing antiperiodic b.c.\
    in the polar direction. b) Wave function amplitude for the
    lowest positive energy state in the $(-,-)$ sector for
    $\Delta_{sj}=0$.  The same parameters are used as in Table I. c)
    Energy eigenvalues of the BdG Hamiltonian plotted against their
    index $n$ for $\Delta_{sj}=0$ (blue) and $\Delta_{sj}=0.2$
    (red). Shaded area corresponds to the bulk SC gap. The in-gap
    states are localized near the edges.}
  \label{fig7}
\end{figure*}

\subsection{Annulus geometry}

To study the model in the annulus geometry we write Hamiltonian
Eq.\ \eqref{h2} in real space and  numerically diagonalize the resulting BdG matrix to find its eigenvalues and eigenstates. Fig.\ \ref{fig7}(a) shows a square lattice cluster that we use to
 approximate an annulus with inner radius $R_1$, outer radius
 $R_2$ and the total number of sites $\sim\pi(R_2^2-R_1^2)$.
 Similar to the strip we use  SC edge term Eq.\ \eqref{h4a} (adapted
 for real-space formulation) to
 gap out the helical edge modes and take $\Delta^{\rm
   edge}=\Delta_{s1}\ (\Delta_{s2})$ for the inner (outer) edge, with
 $\sgn(\Delta_{s1}\Delta_{s2})$ controlling the boundary conditions in
 the radial direction. Antiperiodic b.c.\ in the polar direction is
achieved by inserting a magnetic flux $\Phi_0$ in the hole,
 accompanied by $2\pi$ winding in the SC phase. As explained in Sec.\ III this is
 implemented by attaching a minus sign to  all tunneling and pairing amplitudes
 on bonds that cross the positive $x$ axis, green line in Fig.\
 \ref{fig7}(a). The four possible  boundary
 conditions are labeled by a pair of radial and polar indices
 $(\nu_r,\nu_\varphi)$ that are analogous to  $(\nu_x,\nu_y)$  employed in
 the long strip geometry.
\setlength{\tabcolsep}{6pt} 
\begin{table*}[t]
\centering 
\caption{Parity $P$ and the relative ground state energy
  $\varepsilon^{(\nu_r,\nu_\varphi)}=(E_g^{(\nu_r,\nu_\varphi)}-E_g^{\rm
    avg})/|E_g^{\rm avg}|$ for the
  four ground state sectors in the annulus geometry described in Fig.\
  \ref{fig7}, as a function of increasing on-site disorder $w$. The
  parity is computed from $P=(-1)^{{\rm rank}(V)}$ where $V$ is a
  matrix composed of $v_n$ components of the BdG eigenstate
  $\phi_n=(u_n,v_n)^T$ belonging to positive eigenvalues $E_n$. The
  displayed values of $\varepsilon$ reflect a single disorder
  realization; however, we find very little variation between
  different realizations indicating that for this size the system is
  already in the self-averaging regime.}
\begin{tabular}{c c c c c}
\hline\hline
  $(\nu_r,\nu_\varphi)$ &\  ~\ $(+,+)$\  ~\  ~\ &\  ~\ ~\ $(+,-)$ \  ~\ ~\   &\  ~\ ~\ $(-,+)$ ~\ ~\ &\ 
                                                    ~\ ~\ $(-,-)$ ~\ ~\ \\
\hline
$P$ &            $+$      &  $+$ &  $+$  &  $-$ \\
\hline
$\varepsilon\ (w=0.0)$  & $4.09442 \times 10^{-9}$ & $ -3.72549\times 10^{-9}$
                                                                       & $3.72549\times 10^{-9}$
                                                            &  $-4.0944\times 10^{-9}$ \\
\hline
$\varepsilon\ (w=0.2)$  & $3.94771 \times 10^{-9}$ & $-3.56987\times 10^{-9}$  & $3.56987\times 10^{-9}$
                                                             &
                                                               $-3.94771 \times 10^{-9}$
\\
  \hline
$\varepsilon\ (w=0.4)$  & $3.49598\times 10^{-9}$ & $-3.09333\times 10^{-9}$  & $3.09333\times 10^{-9}$
                                                             &
                                                               $-3.49598\times 10^{-9}$ \\
\hline\hline
\end{tabular}
\label{table2}
\end{table*}

 As shown in Fig.\ \ref{fig7}(b,c) the spectrum of excitations contains
 gapless edge modes when $\Delta_{sj}=0$. Taking non-zero
 $\Delta_{sj}$ produces a gap in the edge mode spectrum. In this
 regime we expect the annulus to behave effectively as a torus. This
 is indeed confirmed by observing four-fold ground state degeneracy
 which obeys the
 characteristic 3:1 parity rule, summarized in Table II. We show
 results for a clean system described by Eqs.\ \eqref{h2} and
 \eqref{h4a} and in the presence of on-site disorder. This is
 implemented by adding
\begin{equation}\label{h21}
\cH_{\rm dis}=\sum_{\br,\sigma}\zeta_\br c_{\br\sigma}^\dagger c_{\br\sigma}
\end{equation}
to the normal-state Hamiltonian, where $\br$ labels the lattice sites
and  $\zeta_\br$ is a random variable drawn from a uniform distribution
between $(-w,w)$. We observe that the fractional energy  splitting
$\varepsilon$ of the
ground-state manifold is small ($\sim 10^{-9}$, just above our
numerical precision) and is essentially independent of the magnitude
of the on-site disorder $w$. This remains true even for strong
disorder with $w=0.4$, comparable to the bulk SC gap.  We conclude
that the pattern of the ground-state degeneracy  expected of a chiral $p$-wave SC on a
torus is robustly present on the annulus.

Table II indicates that on annulus the ground state belonging to the
odd parity sector is $(-,-)$. Intuitively, this can be understood as
follows. In the radial direction we have the same effect as on the
long strip: to complete the cycle electron has to flip spin twice
which adds a $\pi$ Berry phase. In the polar direction there is a
difference between the strip and the annulus rooted in distinct
trajectories along the respective cycle. On the strip the trajectory
is a straight line whereas on the annulus it is a loop encircling
the hole. In a $p_x\pm ip_y$ superconductor a Cooper pair acquires a
$\pm 2\pi$ Berry phase on completing such a planar loop. This can be seen
by thinking about the pair as living on a bond between two sites. Under chiral $p$-wave pairing these bond fields have non-zero relative
phases which add up to $\pm 2\pi$ along any closed, non-intersecting
path. A $2\pi$ phase for a Cooper pair corresponds to a $\pi$ phase
for an electron -- hence, electrons circling a hole experience
effectively antiperiodic boundary conditions. As a result we expect
the 
$(\nu_r,\nu_\varphi)=(-,-)$ sector to provide effectively periodic b.c.\ on the
annulus and hence host the odd-parity ground state.      

\begin{figure}[t]
  \includegraphics[width=8.5cm]{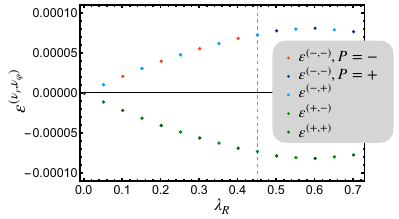}
  \caption{Splitting of the ground-state degeneracy on the annulus in the presence of
    Rashba SOC. Vertical dashed line marks the transition in parity
    $P$ of the $(-,-)$ ground state. The same parameters as used as in
  Table II with $w=0$. Including on-site disorder does not
  significantly change these results.}
  \label{fig8}
\end{figure}

We finally consider the effect of Rashba SOC. Similar to the long
strip geometry we expect $\lambda_R$ to split the ground-state
manifold and eventually remove the 3:1 parity pattern characteristic
of the chiral $p$-wave SC on the  torus. Fig.\ \ref{fig8} shows the
splitting quantified by $\varepsilon^{(\nu_r,\nu_\varphi)}$ as
calculated for the
  annulus with  2392 sites. Rashba SOC is seen to split the manifold into two
  nearly-degenerate sectors labeled by the common $\nu_r$. For small
  $\lambda_R\lesssim 0.2$ the splitting remains negligible compared to the
  excitation gap, the 3:1 parity rule is obeyed and the system on
  annulus can be regarded as  equivalent to the chiral $p$-wave SC on torus.  At
  $\lambda_{Rc}\simeq 0.45$ the edge gap closes marking the transition
  to a phase dominated by SOC where all
  four states belong to the even parity sector. The system can no
  longer be thought of as equivalent to the torus in this regime.

 \section{Discussion}

 \subsection{Summary of results}
 
A 2D quantum system whose spin-up electrons possess  topological
order TO while spin-down electrons possess its time-reversal conjugate,
forming together a combined ${\rm TO}\otimes\overline{\rm TO}$ state, exhibits
ground-state topological 
degeneracy characteristic of the TO on the torus when placed on an
annulus with symmetry-breaking edges. Such an annular geometry is
generically much better suited for experimental investigations than torus. Our
results based on a simple microscopic model can be viewed as
providing microscopic grounding for the ideas introduced recently in
Ref.\ \cite{Wen2024} in the context of  twisted
homobilayer MoTe$_2$ which  may support ${\rm Pf}^\uparrow\otimes\overline {\rm
  Pf}^\downarrow$ topological order  according to the recent experimental report
\cite{Mak2024}.

Alternately, our model is applicable directly to
superconducting altermagnets where it provides a blueprint for probing
topological degeneracy in a new family of materials. The key
theoretical insight is that characteristic  spin-split fermi surfaces
of altermagnets are naturally suited for the emergence of chiral
$p$-wave 
superconductivity which is the leading instability in the presence
of weak attractive interactions \cite{Zhu2023,Sunny2024}. This is a
direct consequence of the  fact that, as
illustrated in Fig.\ \ref{fig2}(a), one must pair same-spin electrons
to form a Cooper pair. For this reason, one generically expects 
metallic altermagnets to exhibit odd-parity SC order when cooled to
sufficiently low temperatures. 

In the strict non-relativistic limit (i.e.\ vanishing Rashba SOC) the
Hamiltonian describing such SC altermagnet remains diagonal in spin
space and obeys the  $\mathbb{Z}_2^\uparrow\times
\mathbb{Z}_2^\downarrow$ spin conservation. This leads to 4 possible
ground states with either 
$p_+$ or $p_-$ orbital pair wave-function  for each spin projection. In the presence of weak Rashba SOC or
spin-singlet SC order spontaneously generated near the edges, the
$p_-^\uparrow\otimes p_+^\downarrow$ helical state becomes
energetically favored, furnishing a concrete 
realization of the ${\rm TO}\otimes\overline{\rm TO}$ state. We
established the ground state degeneracy of this 
state in long strip and annular geometries by an
explicit numerical diagonalization of the relevant Bogoliubov-de
Gennes Hamiltonians. The observed degeneracy
becomes exact in the thermodynamic limit, shows the
3:1 parity pattern expected for the chiral $p$-wave SC on the torus and is
robust with respect to disorder.

When Rashba SOC is present in the
system the protecting
$\mathbb{Z}_2^\uparrow\times\mathbb{Z}_2^\downarrow$ symmetry is
broken down to the global $\mathbb{Z}_2$ parity conservation.  The
ground state degeneracy 
is then split in proportion to the SOC strength $\lambda_R$. For
$\lambda_R$ small compared to the excitation gap the 3:1 parity rule
continues to hold and the system can still be thought of as living
effectively on an `imperfect' or `fuzzy' torus with a weak tunneling between its
opposite walls enabled via SOC. We thus find that, perhaps not
surprisingly,  the topological equivalence between torus and
annulus indicated in Fig.\ \ref{fig1} is exact only when the bulk
of the system respects the protecting symmetry.  This, however,
will be true for any TO, including the 
putative ${\rm Pf}^\uparrow\otimes\overline {\rm Pf}^\downarrow$ phase
in MoTe$_2$. On the annulus, because TO and  $\overline {\rm TO}$ are
overlapping in real space, the topological order becomes symmetry-protected. 

\subsection{Experimental connections}

If twisted bilayer MoTe$_2$ indeed harbor the ${\rm Pf}^\uparrow\otimes\overline {\rm
  Pf}^\downarrow$ phase then annulus-shaped samples could be used to
probe the corresponding ground-state degeneracy as discussed in  Ref.\
\cite{Wen2024}. Conversely, such measurements could be used to shed light on the
nature of the TO in this material in ways that are complementary to the
edge-mode conduction results reported in Ref.\ \cite{Mak2024}. Below
we discuss how this physics would manifest in altermagnet realization
of  ${\rm TO}\otimes\overline{\rm TO}$.

Although no
material realizations of SC altermagnets have been reported to date,
many altermagnet candidates have recently been identified
\cite{Naka2019,Yuan2020,Mazin2021,Spaldin2022,Guo2023,Ostanin2024}   and the
characteristic spin-split electron bands have already been experimentally
observed in some of them
\cite{Krempasky2024,Lee2024,Fedchenko2024,Reimers2024}. Since many of
these materials are good metals one expects superconductivity to occur
in some of them at sufficiently low temperatures. Importantly, the
propensity of altermagnets to form
odd-parity SC order is encoded in their normal-state band structure and
does not require any unusual form of electron-electron interaction;
the ubiquitous phonon
mediated attraction is perfectly adequate. Also, because the magnetism is
compensated, it will not interfere with the onset of superconducting order. 

Assuming a suitable SC altermagnet becomes available in the near
future, how can we use it to experimentally probe the ground-state
degeneracy phenomena discussed in this paper?  Before we attempt to answer this
question it is necessary to highlight an important distinction between
the ground state degeneracy in FQH liquids and superconductors. While
degeneracy 
in both systems is associated with fluxes threading torus
holes, in FQH the gauge field is emergent and as such exists only
within the material itself. The degeneracy is intrinsic in that
the relevant many-body Hamiltonian fully captures the gauge field
dynamics, and, when diagonalized,
exhibits the requisite number of degenerate ground states in its
spectrum. In superconductors, by contrast, the degeneracy is
associated with the real electromagnetic field which exists outside
the material and enters  the BdG Hamiltonian as an external
parameter, see e.g.\ Eq.\ \eqref{h12}. It would
become a true degeneracy if the EM field were treated as
dynamical. However, we know that at energy and length scales
that are at play in solid-state systems quantum fluctuations of
the magnetic flux through the SC ring are negligible.  Hence, the toroidal
ground state degeneracy in a superconductor is best regarded as a statement of
energetic equivalence between states containing even vs.\ odd number of
SC flux quanta.
\begin{figure}[t]
  \includegraphics[width=7.0cm]{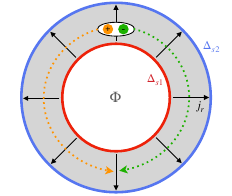}
  \caption{Schematic setup for probing the ground state degeneracy in  the
    $p_-^\uparrow\otimes p_+^\downarrow$ superconducting
    ring. Magnetic flux $\Phi$ is used to control index $\nu_\varphi$ while the
    radial current density $j_r$ can be applied to promote
    vortex/antivortex unbinding and recombination. As explained in the
  text such events switch between the two $\nu_r$ sectors.}
  \label{fig9}
\end{figure}

When viewed in this light the classic Little-Parks flux quantization
experiment \cite{Doll1961,Little1962} can be regarded as experimental
confirmation of sorts of 
the expected energetic equivalence in conventional spin-singlet
superconductors. The question then arises: Can we use the
torus/annulus equivalence to go further and probe the degeneracy
specific to the chiral $p$-wave superconductor? In practice this would
entail testing the 3:1 parity rule that distinguishes the chiral
$p$-wave state. In the remainder of this subsection we outline 
possible strategies to achieve this.

At a minimum, to probe the topological degeneracy, one needs a way to
switch between the energy minima controlled by the boundary conditions
labeled by the pair of indices $(\nu_r,\nu_\varphi)$.  
Inspection of Fig.\ \ref{fig7}(a) and Table II suggests the following
approach. The $\nu_\varphi$ index can be controlled straightforwardly
by adjusting the applied magnetic  field $B$ 
perpendicular to the plane of the annulus. For practical reasons it is
best to employ a ring geometry with a large hole (i.e.\ $R_1$ only slightly
smaller than $R_2$) such that the flux $\Phi_0=\pi R_1^2B$ through the
hole corresponds to a sufficiently weak field permeating the
superconductor as can be neglected. Then one could switch between the
two $\nu_\varphi$ sectors by simply changing the applied field.

The two
$\nu_r$ sectors are controlled by the relative sign of
$\Delta_{s1}$ and $\Delta_{s2}$. Switching between them can
be achieved by nucleating a vortex-antivortex pair in one spin sector, dragging its
constituents around the hole as indicated in  Fig.\ \ref{fig9}, and
then annihilating the pair. Vortex-antivortex pairs occur naturally as
quantum or thermal fluctuations in a 2D superconductor. Deep
in the SC phase such pairs  are bound and the probability of a pair
dissociating and recombining after its constituents have travelled along the opposite
branches of the annulus is negligible. Imagine, however, that we
attach electrical contacts to the inner and 
outer edges of the annulus and drive current $J_r$ between them. Ideally,
the current density will be uniform around the ring with the current
flowing in the radial direction as indicated in Fig.\ \ref{fig9}, but
this is not essential. In the presence of the
radial current the members of a spontaneously nucleated
vortex-antivortex pair  will feel Magnus force that is
perpendicular to the current direction  and opposite for the vortex
and the antivortex. 
For a sufficiently large applied current one
expects the probability of the above encircling process to become
significant. In essence, the applied current will tear apart a
vortex-antivortex pair that has become sufficiently large such that
the Magnus force overcomes their attraction. It will also  assist
their journey around the hole. Importantly, such a process is
analogous to a phase slip which 
advances the phase difference between the inner and the outer edge by
$2\pi$. If a single such event occurs in one spin species, the net effect
dictated by Eq.\ \eqref{h19} is precisely to change the relative sign
between $\Delta_{s1}$ and $\Delta_{s2}$.

We conclude that, at least in principle, it is possible to switch
between the 4 ground state sectors of the annulus by adjusting the
applied magnetic field $B$ and
the radial current $J_r$, as discussed above. Now suppose we perform a Little-Parks type
experiment by increasing $J_r$ close to its critical value. In this
regime the resistance will be dominated by phase slips and we can
think of the annulus as flipping between $\nu_r=\pm$ sectors. According
to Table II, for zero flux $\Phi$  through the hole (corresponding to the
$\nu_\varphi=+$ sector), both $\nu_r$ ground states belong to the same
even parity subspace.  For $\Phi=\Phi_0$, the system will be in the
$\nu_\varphi=-$ sector. In this case, flipping $\nu_r$ will cause the
system to switch between even and odd parity ground states. If the
ring is electronically isolated such that the parity $P$ is conserved
then switching from the $(-,+)$ sector to $(-,-)$ will drive the
system to an excited state, with one fermionic quasiparticle above the
odd-parity ground state. Hence, one would expect to see an energetic
difference in Little-Parks oscillations between zero-flux and
$\Phi=\Phi_0$ cases. If the parity is not conserved then there will be
no energy difference; in this case one could employ a direct parity
detection scheme \cite{Devoret1993,Pekola2019} to identify the
$\nu_\varphi=\pm$ sectors.

\subsection{Outlook}

Theoretical developments in Ref.\ \cite{Wen2024} and in this work
suggest that topological degeneracy, first conceptualized in works of
Haldane, Wen and others more than 30 years ago
\cite{Haldane1985,Wen1990,Wen1991}, could be 
experimentally tested 
in a class of systems that support two copies of time-reversed
topological order  ${\rm TO}\otimes\overline{\rm TO}$. In this study
we performed a detailed analysis of a specific example furnished by a
$p_-^\uparrow\otimes p_+^\downarrow$ superconducting phase that we expect to naturally
form in  metallic altermagnets at sufficiently low temperatures. By
virtue of the Read-Green correspondence \cite{Read2000} this phase shares
key topological properties with the ${\rm Pf}^\uparrow\otimes\overline {\rm
  Pf}^\downarrow$ fractional quantum Hall liquid that may be realized
in twisted homobilayers of MoTe$_2$.  The chief advantage of
considering  the $p_-^\uparrow\otimes p_+^\downarrow$ superconducting
example lies in its tractability: a treatment at the level of the conventional mean-field BCS
theory allowed us to explicitly evaluate ground state energies, the
corresponding parity eigenvalues, and confirm in detail the expected
ground state degeneracies for various sample geometries, including the
experimentally accessible annulus.

Our main conclusion based on these
results is that a SC altermagnet fabricated in a ring geometry Fig.\
\ref{fig7}(a) could indeed be used to probe the topological
degeneracy. A key requirement is that the material respects the 
$\mathbb{Z}_2^\uparrow\times\mathbb{Z}_2^\downarrow$
spin-conservation symmetry in the bulk (i.e.\ away from the edges). In
practice this means negligible Rashba SOC as well as any perturbations that
flip electron spin, such as magnetic impurities. We showed that ground
state degeneracy is robust to non-magnetic disorder but spin-flip
inducing perturbations are detrimental in that they split the
degeneracy in proportion to their strength. Nevertheless, some key signatures, such as the
characteristic 3:1 parity property, persist up to a critical strength
of  spin-flip
perturbations. We note that this requirement is generic and applies to
any ${\rm TO}\otimes\overline{\rm TO}$.

We conclude by listing possible broader impacts of this work.
A ring with $p_-^\uparrow\otimes p_+^\downarrow$ superconducting order
depicted in Fig.\ \ref{fig7}(a) could be used as a protected qubit.
We envision the two logical states encoded as two ground states labeled
by $\nu_r=\pm$ with fixed $\nu_\varphi$ set by the external magnetic flux
$\Phi$. The qubit is protected because the information is encoded {\em
  non-locally} in $\sgn(\Delta_{s1}\Delta_{s2})$ and measuring one
$\Delta_{sj}$ alone does not provide any information on the state.
Instead, to read out the qubit, one must perform a measurement of the
relative phase between $\Delta_{s1}$ and $\Delta_{s2}$ and this is
necessarily a non-local measurement given that inner and outer
edges are macroscopically separated. In the $\nu_\varphi=-$ sector the
product $\sgn(\Delta_{s1}\Delta_{s2})$ encodes the ring electron
parity $P$, in a way reminiscent of two Majorana zero modes localized
e.g.\ at the ends of a quantum wire. In the present case, however, the
underlying system is fully gapped and hence, presumably, more robustly
immune to decoherence.  

Another
important realization is the fact that metallic altermagnets are
generically expected to harbor odd-parity superconducting orders, with
chiral $p$-wave being the most natural in the square lattice and $f$-wave
in hexagonal crystals. This has been first noted in Ref.\
\cite{Zhu2023} and our work \cite{Sunny2024} confirms this expectation. Such an exotic
superconductivity has been long sought by the materials community  but
never convincingly demonstrated in any material to date. Another
logical possibility, not considered in this paper,  is
spin-singlet pairing at non-zero Cooper pair  momentum 
(the Fulde-Ferrel-Larkin-Ovchinikov type state) discussed in Ref.\
\cite{Chakraborty2024}. Such a state is also exotic and rare, especially in zero
applied magnetic field. Hence, comprehensive search for superconducting
instabilities in metallic altermagnets constitutes  a promising
direction for future research, likely to yield exciting discoveries.

\section*{Acknowledgments}

We are grateful to Ching-Kai Chiu, Niclas Heinsdorf, Masaki Oshikawa,
Adarsh Patri, Andrew Potter, Nicholas Read, Rui Wen and  Ashvin
Vishwanath for illuminating discussions and correspondence. This
research was supported in part by NSERC, CIFAR, and the Canada First
Research Excellence Fund, Quantum Materials and Future Technologies
Program. 
\begin{figure}[b]
  \includegraphics[width=8.5cm]{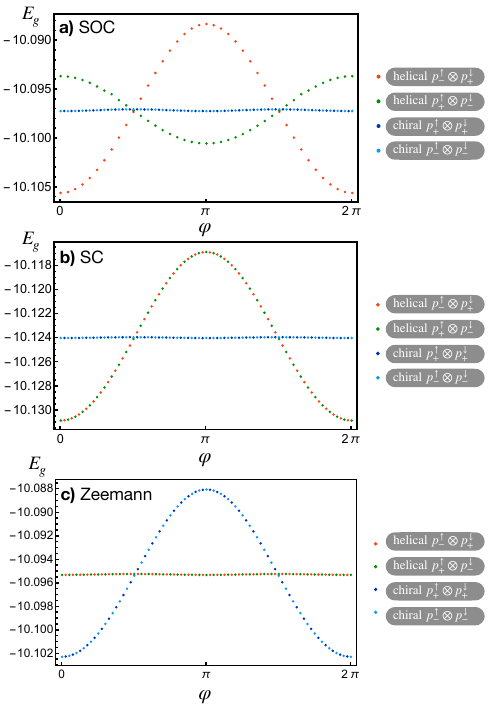}
  \caption{Ground state energy as a function of the relative phase
    $\varphi$ in the presence of (a) Rashba SOC,  (b) spin-singlet
    SC order parameter and (c) in-plane external magnetic field
    $B_x$. We use $\eta=0.2$, 
  $\mu=-2.1$, $\Delta_\sigma=0.5$ and $(\lambda_R,\Delta_s,B_x)=(0.1,0.0,0.0)$
in (a) $(0.0,0.2,0.0)$ in (b)  and $(0.0,0.0,0.2)$ in (c).}
  \label{figA1}
\end{figure}
%

\appendix	

\section{Effect of Rashba SOC and spin-singlet SC}\label{app1}

We wish to understand the effect of adding a weak Rashba SOC or a spin-singlet
SC order parameter component on 4 degenerate ground states discussed in Sec.\
IIB. The former arises generically when a sample is placed on a
substrate which breaks the inversion symmetry of the model. The latter
can then arise spontaneously from attractive interactions in the
sample or can be induced externally by a proximity effect with a
conventional spin-singlet superconductor.

To study the energetics of the model in the presence of these
perturbation we compute the many-body ground state energy given for
the Hamiltonian Eq. \eqref{h2} as
\begin{equation}\label{a1}
 E_g(\varphi)= -\sum_\bk E_{\bk\mu}(\varphi),
\end{equation}
where $E_{\bk\mu}$ are the positive eigenvalues of the BdG matrix
$H_\bk$. This energy will depend on the relative phase $\varphi$ of
the two SC order parameters which are uncoupled in the unperturbed
system. Specifically, we assume $\Delta_\uparrow=\Delta_p$ and 
$\Delta_\downarrow=e^{i\varphi}\Delta_p$ with $\Delta_p$ real
positive.

The result in the presence of Rashba SOC is given in Fig.\
\ref{figA1}(a). It shows that nonzero $\lambda_R$ favors the helical
$p_-^\uparrow\otimes p_+^\downarrow$ state and the absolute minimum
occurs at $\varphi=0$.

To study the effect of spin-singlet SC component we generalize the
pair-field matrix entering the BdG  Hamiltonian as follows,
\begin{equation}\label{a2}
  \hat\Delta_\bk=
  \begin{pmatrix}
    \Delta_{\bk\uparrow} & \Delta_s \\
    \Delta_s &  \Delta_{\bk\downarrow}
      \end{pmatrix},
\end{equation}
where $\Delta_s$ is the spin-singlet amplitude. The resulting ground
state energy is shown in  Fig.\ \ref{figA1}(b). Once again the absolute
minimum occurs at $\varphi=0$ but in this case the two helical states
remain degenerate while the chiral states are pushed to higher
energy.

Panel (c) finally shows the effect of in-plane applied magnetic field
$B_x$. This has the opposite effect to the two $\cT$-respecting
perturbations and selects the two chiral phases as ground states.

We thus conclude that both $\cT$-respecting perturbations favor the helical states in
accord with the intuitive reasoning presented in the main text and
with Ref.\ \cite{Zhu2023}.  When both Rashba SOC and with non-zero
$\Delta_s$ are present, there is a single non-degenerate ground state characterized as $p_-^\uparrow\otimes p_+^\downarrow$.

\bibliography{pip}

\begin{thebibliography}{48}%
\makeatletter
\providecommand \@ifxundefined [1]{%
 \@ifx{#1\undefined}
}%
\providecommand \@ifnum [1]{%
 \ifnum #1\expandafter \@firstoftwo
 \else \expandafter \@secondoftwo
 \fi
}%
\providecommand \@ifx [1]{%
 \ifx #1\expandafter \@firstoftwo
 \else \expandafter \@secondoftwo
 \fi
}%
\providecommand \natexlab [1]{#1}%
\providecommand \enquote  [1]{``#1''}%
\providecommand \bibnamefont  [1]{#1}%
\providecommand \bibfnamefont [1]{#1}%
\providecommand \citenamefont [1]{#1}%
\providecommand \href@noop [0]{\@secondoftwo}%
\providecommand \href [0]{\begingroup \@sanitize@url \@href}%
\providecommand \@href[1]{\@@startlink{#1}\@@href}%
\providecommand \@@href[1]{\endgroup#1\@@endlink}%
\providecommand \@sanitize@url [0]{\catcode `\\12\catcode `\$12\catcode
  `\&12\catcode `\#12\catcode `\^12\catcode `\_12\catcode `\%12\relax}%
\providecommand \@@startlink[1]{}%
\providecommand \@@endlink[0]{}%
\providecommand \url  [0]{\begingroup\@sanitize@url \@url }%
\providecommand \@url [1]{\endgroup\@href {#1}{\urlprefix }}%
\providecommand \urlprefix  [0]{URL }%
\providecommand \Eprint [0]{\href }%
\providecommand \doibase [0]{http://dx.doi.org/}%
\providecommand \selectlanguage [0]{\@gobble}%
\providecommand \bibinfo  [0]{\@secondoftwo}%
\providecommand \bibfield  [0]{\@secondoftwo}%
\providecommand \translation [1]{[#1]}%
\providecommand \BibitemOpen [0]{}%
\providecommand \bibitemStop [0]{}%
\providecommand \bibitemNoStop [0]{.\EOS\space}%
\providecommand \EOS [0]{\spacefactor3000\relax}%
\providecommand \BibitemShut  [1]{\csname bibitem#1\endcsname}%
\let\auto@bib@innerbib\@empty
\bibitem [{\citenamefont {Haldane}\ and\ \citenamefont
  {Rezayi}(1985)}]{Haldane1985}%
  \BibitemOpen
  \bibfield  {author} {\bibinfo {author} {\bibfnamefont {F.~D.~M.}\
  \bibnamefont {Haldane}}\ and\ \bibinfo {author} {\bibfnamefont {E.~H.}\
  \bibnamefont {Rezayi}},\ }\href {\doibase 10.1103/PhysRevB.31.2529}
  {\bibfield  {journal} {\bibinfo  {journal} {Phys. Rev. B}\ }\textbf {\bibinfo
  {volume} {31}},\ \bibinfo {pages} {2529} (\bibinfo {year}
  {1985})}\BibitemShut {NoStop}%
\bibitem [{\citenamefont {Wen}\ and\ \citenamefont {Niu}(1990)}]{Wen1990}%
  \BibitemOpen
  \bibfield  {author} {\bibinfo {author} {\bibfnamefont {X.~G.}\ \bibnamefont
  {Wen}}\ and\ \bibinfo {author} {\bibfnamefont {Q.}~\bibnamefont {Niu}},\
  }\href {\doibase 10.1103/PhysRevB.41.9377} {\bibfield  {journal} {\bibinfo
  {journal} {Phys. Rev. B}\ }\textbf {\bibinfo {volume} {41}},\ \bibinfo
  {pages} {9377} (\bibinfo {year} {1990})}\BibitemShut {NoStop}%
\bibitem [{\citenamefont {Wen}(1991{\natexlab{a}})}]{Wen1991}%
  \BibitemOpen
  \bibfield  {author} {\bibinfo {author} {\bibfnamefont {X.-G.}\ \bibnamefont
  {Wen}},\ }\href {\doibase 10.1142/S0217979291001541} {\bibfield  {journal}
  {\bibinfo  {journal} {International Journal of Modern Physics B}\ }\textbf
  {\bibinfo {volume} {05}},\ \bibinfo {pages} {1641} (\bibinfo {year}
  {1991}{\natexlab{a}})}\BibitemShut {NoStop}%
\bibitem [{\citenamefont {Nakamura}\ \emph {et~al.}(2020)\citenamefont
  {Nakamura}, \citenamefont {Liang}, \citenamefont {Gardner},\ and\
  \citenamefont {Manfra}}]{Manfra2020}%
  \BibitemOpen
  \bibfield  {author} {\bibinfo {author} {\bibfnamefont {J.}~\bibnamefont
  {Nakamura}}, \bibinfo {author} {\bibfnamefont {S.}~\bibnamefont {Liang}},
  \bibinfo {author} {\bibfnamefont {G.~C.}\ \bibnamefont {Gardner}}, \ and\
  \bibinfo {author} {\bibfnamefont {M.~J.}\ \bibnamefont {Manfra}},\ }\href
  {\doibase 10.1038/s41567-020-1019-1} {\bibfield  {journal} {\bibinfo
  {journal} {Nature Physics}\ }\textbf {\bibinfo {volume} {16}},\ \bibinfo
  {pages} {931} (\bibinfo {year} {2020})}\BibitemShut {NoStop}%
\bibitem [{\citenamefont {Kundu}\ \emph {et~al.}(2023)\citenamefont {Kundu},
  \citenamefont {Biswas}, \citenamefont {Ofek}, \citenamefont {Umansky},\ and\
  \citenamefont {Heiblum}}]{Heiblum2023}%
  \BibitemOpen
  \bibfield  {author} {\bibinfo {author} {\bibfnamefont {H.~K.}\ \bibnamefont
  {Kundu}}, \bibinfo {author} {\bibfnamefont {S.}~\bibnamefont {Biswas}},
  \bibinfo {author} {\bibfnamefont {N.}~\bibnamefont {Ofek}}, \bibinfo {author}
  {\bibfnamefont {V.}~\bibnamefont {Umansky}}, \ and\ \bibinfo {author}
  {\bibfnamefont {M.}~\bibnamefont {Heiblum}},\ }\href {\doibase
  10.1038/s41567-022-01899-z} {\bibfield  {journal} {\bibinfo  {journal}
  {Nature Physics}\ }\textbf {\bibinfo {volume} {19}},\ \bibinfo {pages} {515}
  (\bibinfo {year} {2023})}\BibitemShut {NoStop}%
\bibitem [{\citenamefont {Leinaas}\ and\ \citenamefont
  {Myrheim}(1977)}]{Leinaas1977}%
  \BibitemOpen
  \bibfield  {author} {\bibinfo {author} {\bibfnamefont {J.}~\bibnamefont
  {Leinaas}}\ and\ \bibinfo {author} {\bibfnamefont {J.}~\bibnamefont
  {Myrheim}},\ }\href@noop {} {\bibfield  {journal} {\bibinfo  {journal} {Il
  nuovo cimento}\ }\textbf {\bibinfo {volume} {37}},\ \bibinfo {pages} {132}
  (\bibinfo {year} {1977})}\BibitemShut {NoStop}%
\bibitem [{\citenamefont {Wilczek}(1982)}]{Wilczek1982}%
  \BibitemOpen
  \bibfield  {author} {\bibinfo {author} {\bibfnamefont {F.}~\bibnamefont
  {Wilczek}},\ }\href@noop {} {\bibfield  {journal} {\bibinfo  {journal}
  {Physical review letters}\ }\textbf {\bibinfo {volume} {49}},\ \bibinfo
  {pages} {957} (\bibinfo {year} {1982})}\BibitemShut {NoStop}%
\bibitem [{\citenamefont {Wen}(1991{\natexlab{b}})}]{Wen1991b}%
  \BibitemOpen
  \bibfield  {author} {\bibinfo {author} {\bibfnamefont {X.~G.}\ \bibnamefont
  {Wen}},\ }\href {\doibase 10.1103/PhysRevLett.66.802} {\bibfield  {journal}
  {\bibinfo  {journal} {Phys. Rev. Lett.}\ }\textbf {\bibinfo {volume} {66}},\
  \bibinfo {pages} {802} (\bibinfo {year} {1991}{\natexlab{b}})}\BibitemShut
  {NoStop}%
\bibitem [{\citenamefont {Greiter}\ and\ \citenamefont
  {Wilczek}(2024)}]{Greiter2024}%
  \BibitemOpen
  \bibfield  {author} {\bibinfo {author} {\bibfnamefont {M.}~\bibnamefont
  {Greiter}}\ and\ \bibinfo {author} {\bibfnamefont {F.}~\bibnamefont
  {Wilczek}},\ }\href {\doibase
  https://doi.org/10.1146/annurev-conmatphys-040423-014045} {\bibfield
  {journal} {\bibinfo  {journal} {Annual Review of Condensed Matter Physics}\
  }\textbf {\bibinfo {volume} {15}},\ \bibinfo {pages} {131} (\bibinfo {year}
  {2024})}\BibitemShut {NoStop}%
\bibitem [{\citenamefont {Verlinde}(1988)}]{Verlinde1988}%
  \BibitemOpen
  \bibfield  {author} {\bibinfo {author} {\bibfnamefont {E.}~\bibnamefont
  {Verlinde}},\ }\href {\doibase https://doi.org/10.1016/0550-3213(88)90603-7}
  {\bibfield  {journal} {\bibinfo  {journal} {Nuclear Physics B}\ }\textbf
  {\bibinfo {volume} {300}},\ \bibinfo {pages} {360} (\bibinfo {year}
  {1988})}\BibitemShut {NoStop}%
\bibitem [{\citenamefont {Einarsson}(1990)}]{Einarsson1990}%
  \BibitemOpen
  \bibfield  {author} {\bibinfo {author} {\bibfnamefont {T.}~\bibnamefont
  {Einarsson}},\ }\href {\doibase 10.1103/PhysRevLett.64.1995} {\bibfield
  {journal} {\bibinfo  {journal} {Phys. Rev. Lett.}\ }\textbf {\bibinfo
  {volume} {64}},\ \bibinfo {pages} {1995} (\bibinfo {year}
  {1990})}\BibitemShut {NoStop}%
\bibitem [{\citenamefont {Oshikawa}\ and\ \citenamefont
  {Senthil}(2006)}]{Oshikawa2006}%
  \BibitemOpen
  \bibfield  {author} {\bibinfo {author} {\bibfnamefont {M.}~\bibnamefont
  {Oshikawa}}\ and\ \bibinfo {author} {\bibfnamefont {T.}~\bibnamefont
  {Senthil}},\ }\href {\doibase 10.1103/PhysRevLett.96.060601} {\bibfield
  {journal} {\bibinfo  {journal} {Phys. Rev. Lett.}\ }\textbf {\bibinfo
  {volume} {96}},\ \bibinfo {pages} {060601} (\bibinfo {year}
  {2006})}\BibitemShut {NoStop}%
\bibitem [{\citenamefont {Oshikawa}\ \emph {et~al.}(2007)\citenamefont
  {Oshikawa}, \citenamefont {Kim}, \citenamefont {Shtengel}, \citenamefont
  {Nayak},\ and\ \citenamefont {Tewari}}]{Oshikawa2007}%
  \BibitemOpen
  \bibfield  {author} {\bibinfo {author} {\bibfnamefont {M.}~\bibnamefont
  {Oshikawa}}, \bibinfo {author} {\bibfnamefont {Y.~B.}\ \bibnamefont {Kim}},
  \bibinfo {author} {\bibfnamefont {K.}~\bibnamefont {Shtengel}}, \bibinfo
  {author} {\bibfnamefont {C.}~\bibnamefont {Nayak}}, \ and\ \bibinfo {author}
  {\bibfnamefont {S.}~\bibnamefont {Tewari}},\ }\href {\doibase
  https://doi.org/10.1016/j.aop.2006.08.001} {\bibfield  {journal} {\bibinfo
  {journal} {Annals of Physics}\ }\textbf {\bibinfo {volume} {322}},\ \bibinfo
  {pages} {1477} (\bibinfo {year} {2007})}\BibitemShut {NoStop}%
\bibitem [{\citenamefont {Kitaev}(2003)}]{Kitaev2003}%
  \BibitemOpen
  \bibfield  {author} {\bibinfo {author} {\bibfnamefont {A.}~\bibnamefont
  {Kitaev}},\ }\href {\doibase https://doi.org/10.1016/S0003-4916(02)00018-0}
  {\bibfield  {journal} {\bibinfo  {journal} {Annals of Physics}\ }\textbf
  {\bibinfo {volume} {303}},\ \bibinfo {pages} {2} (\bibinfo {year}
  {2003})}\BibitemShut {NoStop}%
\bibitem [{\citenamefont {Nayak}\ \emph {et~al.}(2008)\citenamefont {Nayak},
  \citenamefont {Simon}, \citenamefont {Stern}, \citenamefont {Freedman},\ and\
  \citenamefont {Das~Sarma}}]{Nayak2008}%
  \BibitemOpen
  \bibfield  {author} {\bibinfo {author} {\bibfnamefont {C.}~\bibnamefont
  {Nayak}}, \bibinfo {author} {\bibfnamefont {S.~H.}\ \bibnamefont {Simon}},
  \bibinfo {author} {\bibfnamefont {A.}~\bibnamefont {Stern}}, \bibinfo
  {author} {\bibfnamefont {M.}~\bibnamefont {Freedman}}, \ and\ \bibinfo
  {author} {\bibfnamefont {S.}~\bibnamefont {Das~Sarma}},\ }\href {\doibase
  10.1103/RevModPhys.80.1083} {\bibfield  {journal} {\bibinfo  {journal} {Rev.
  Mod. Phys.}\ }\textbf {\bibinfo {volume} {80}},\ \bibinfo {pages} {1083}
  (\bibinfo {year} {2008})}\BibitemShut {NoStop}%
\bibitem [{\citenamefont {Wen}\ and\ \citenamefont {Potter}(2024)}]{Wen2024}%
  \BibitemOpen
  \bibfield  {author} {\bibinfo {author} {\bibfnamefont {R.}~\bibnamefont
  {Wen}}\ and\ \bibinfo {author} {\bibfnamefont {A.~C.}\ \bibnamefont
  {Potter}},\ }\href {https://arxiv.org/abs/2407.03401} {\enquote {\bibinfo
  {title} {Cheshire qudits from fractional quantum spin hall states in twisted
  mote$_2$},}\ } (\bibinfo {year} {2024}),\ \Eprint
  {http://arxiv.org/abs/2407.03401} {arXiv:2407.03401 [cond-mat.str-el]}
  \BibitemShut {NoStop}%
\bibitem [{\citenamefont {Kang}\ \emph {et~al.}(2024)\citenamefont {Kang},
  \citenamefont {Shen}, \citenamefont {Qiu}, \citenamefont {Zeng},
  \citenamefont {Xia}, \citenamefont {Watanabe}, \citenamefont {Taniguchi},
  \citenamefont {Shan},\ and\ \citenamefont {Mak}}]{Mak2024}%
  \BibitemOpen
  \bibfield  {author} {\bibinfo {author} {\bibfnamefont {K.}~\bibnamefont
  {Kang}}, \bibinfo {author} {\bibfnamefont {B.}~\bibnamefont {Shen}}, \bibinfo
  {author} {\bibfnamefont {Y.}~\bibnamefont {Qiu}}, \bibinfo {author}
  {\bibfnamefont {Y.}~\bibnamefont {Zeng}}, \bibinfo {author} {\bibfnamefont
  {Z.}~\bibnamefont {Xia}}, \bibinfo {author} {\bibfnamefont {K.}~\bibnamefont
  {Watanabe}}, \bibinfo {author} {\bibfnamefont {T.}~\bibnamefont {Taniguchi}},
  \bibinfo {author} {\bibfnamefont {J.}~\bibnamefont {Shan}}, \ and\ \bibinfo
  {author} {\bibfnamefont {K.~F.}\ \bibnamefont {Mak}},\ }\href {\doibase
  10.1038/s41586-024-07214-5} {\bibfield  {journal} {\bibinfo  {journal}
  {Nature}\ }\textbf {\bibinfo {volume} {628}},\ \bibinfo {pages} {522}
  (\bibinfo {year} {2024})}\BibitemShut {NoStop}%
\bibitem [{\citenamefont {Moore}\ and\ \citenamefont {Read}(1991)}]{Moore1991}%
  \BibitemOpen
  \bibfield  {author} {\bibinfo {author} {\bibfnamefont {G.}~\bibnamefont
  {Moore}}\ and\ \bibinfo {author} {\bibfnamefont {N.}~\bibnamefont {Read}},\
  }\href {\doibase https://doi.org/10.1016/0550-3213(91)90407-O} {\bibfield
  {journal} {\bibinfo  {journal} {Nuclear Physics B}\ }\textbf {\bibinfo
  {volume} {360}},\ \bibinfo {pages} {362} (\bibinfo {year}
  {1991})}\BibitemShut {NoStop}%
\bibitem [{\citenamefont {Reddy}\ \emph {et~al.}(2024)\citenamefont {Reddy},
  \citenamefont {Paul}, \citenamefont {Abouelkomsan},\ and\ \citenamefont
  {Fu}}]{Reddy2024}%
  \BibitemOpen
  \bibfield  {author} {\bibinfo {author} {\bibfnamefont {A.~P.}\ \bibnamefont
  {Reddy}}, \bibinfo {author} {\bibfnamefont {N.}~\bibnamefont {Paul}},
  \bibinfo {author} {\bibfnamefont {A.}~\bibnamefont {Abouelkomsan}}, \ and\
  \bibinfo {author} {\bibfnamefont {L.}~\bibnamefont {Fu}},\ }\href {\doibase
  10.1103/PhysRevLett.133.166503} {\bibfield  {journal} {\bibinfo  {journal}
  {Phys. Rev. Lett.}\ }\textbf {\bibinfo {volume} {133}},\ \bibinfo {pages}
  {166503} (\bibinfo {year} {2024})}\BibitemShut {NoStop}%
\bibitem [{\citenamefont {Ahn}\ \emph {et~al.}(2024)\citenamefont {Ahn},
  \citenamefont {Lee}, \citenamefont {Yananose}, \citenamefont {Kim},\ and\
  \citenamefont {Cho}}]{Ahn2024}%
  \BibitemOpen
  \bibfield  {author} {\bibinfo {author} {\bibfnamefont {C.-E.}\ \bibnamefont
  {Ahn}}, \bibinfo {author} {\bibfnamefont {W.}~\bibnamefont {Lee}}, \bibinfo
  {author} {\bibfnamefont {K.}~\bibnamefont {Yananose}}, \bibinfo {author}
  {\bibfnamefont {Y.}~\bibnamefont {Kim}}, \ and\ \bibinfo {author}
  {\bibfnamefont {G.~Y.}\ \bibnamefont {Cho}},\ }\href {\doibase
  10.1103/PhysRevB.110.L161109} {\bibfield  {journal} {\bibinfo  {journal}
  {Phys. Rev. B}\ }\textbf {\bibinfo {volume} {110}},\ \bibinfo {pages}
  {L161109} (\bibinfo {year} {2024})}\BibitemShut {NoStop}%
\bibitem [{\citenamefont {Read}\ and\ \citenamefont {Green}(2000)}]{Read2000}%
  \BibitemOpen
  \bibfield  {author} {\bibinfo {author} {\bibfnamefont {N.}~\bibnamefont
  {Read}}\ and\ \bibinfo {author} {\bibfnamefont {D.}~\bibnamefont {Green}},\
  }\href {\doibase 10.1103/PhysRevB.61.10267} {\bibfield  {journal} {\bibinfo
  {journal} {Phys. Rev. B}\ }\textbf {\bibinfo {volume} {61}},\ \bibinfo
  {pages} {10267} (\bibinfo {year} {2000})}\BibitemShut {NoStop}%
\bibitem [{\citenamefont {Hayami}\ \emph {et~al.}(2019)\citenamefont {Hayami},
  \citenamefont {Yanagi},\ and\ \citenamefont {Kusunose}}]{Hayami2020}%
  \BibitemOpen
  \bibfield  {author} {\bibinfo {author} {\bibfnamefont {S.}~\bibnamefont
  {Hayami}}, \bibinfo {author} {\bibfnamefont {Y.}~\bibnamefont {Yanagi}}, \
  and\ \bibinfo {author} {\bibfnamefont {H.}~\bibnamefont {Kusunose}},\ }\href
  {\doibase 10.7566/JPSJ.88.123702} {\bibfield  {journal} {\bibinfo  {journal}
  {Journal of the Physical Society of Japan}\ }\textbf {\bibinfo {volume}
  {88}},\ \bibinfo {pages} {123702} (\bibinfo {year} {2019})}\BibitemShut
  {NoStop}%
\bibitem [{\citenamefont {\ifmmode~\check{S}\else \v{S}\fi{}mejkal}\ \emph
  {et~al.}(2022{\natexlab{a}})\citenamefont {\ifmmode~\check{S}\else
  \v{S}\fi{}mejkal}, \citenamefont {Sinova},\ and\ \citenamefont
  {Jungwirth}}]{Smejkal2022a}%
  \BibitemOpen
  \bibfield  {author} {\bibinfo {author} {\bibfnamefont {L.}~\bibnamefont
  {\ifmmode~\check{S}\else \v{S}\fi{}mejkal}}, \bibinfo {author} {\bibfnamefont
  {J.}~\bibnamefont {Sinova}}, \ and\ \bibinfo {author} {\bibfnamefont
  {T.}~\bibnamefont {Jungwirth}},\ }\href {\doibase 10.1103/PhysRevX.12.031042}
  {\bibfield  {journal} {\bibinfo  {journal} {Phys. Rev. X}\ }\textbf {\bibinfo
  {volume} {12}},\ \bibinfo {pages} {031042} (\bibinfo {year}
  {2022}{\natexlab{a}})}\BibitemShut {NoStop}%
\bibitem [{\citenamefont {\ifmmode~\check{S}\else \v{S}\fi{}mejkal}\ \emph
  {et~al.}(2022{\natexlab{b}})\citenamefont {\ifmmode~\check{S}\else
  \v{S}\fi{}mejkal}, \citenamefont {Sinova},\ and\ \citenamefont
  {Jungwirth}}]{Smejkal2022b}%
  \BibitemOpen
  \bibfield  {author} {\bibinfo {author} {\bibfnamefont {L.}~\bibnamefont
  {\ifmmode~\check{S}\else \v{S}\fi{}mejkal}}, \bibinfo {author} {\bibfnamefont
  {J.}~\bibnamefont {Sinova}}, \ and\ \bibinfo {author} {\bibfnamefont
  {T.}~\bibnamefont {Jungwirth}},\ }\href {\doibase 10.1103/PhysRevX.12.040501}
  {\bibfield  {journal} {\bibinfo  {journal} {Phys. Rev. X}\ }\textbf {\bibinfo
  {volume} {12}},\ \bibinfo {pages} {040501} (\bibinfo {year}
  {2022}{\natexlab{b}})}\BibitemShut {NoStop}%
\bibitem [{\citenamefont {Mazin}(2022)}]{Mazin2022}%
  \BibitemOpen
  \bibfield  {author} {\bibinfo {author} {\bibfnamefont {I.}~\bibnamefont
  {Mazin}} (\bibinfo {collaboration} {The PRX Editors}),\ }\href {\doibase
  10.1103/PhysRevX.12.040002} {\bibfield  {journal} {\bibinfo  {journal} {Phys.
  Rev. X}\ }\textbf {\bibinfo {volume} {12}},\ \bibinfo {pages} {040002}
  (\bibinfo {year} {2022})}\BibitemShut {NoStop}%
\bibitem [{\citenamefont {Zhu}\ \emph {et~al.}(2024)\citenamefont {Zhu},
  \citenamefont {Chen}, \citenamefont {Liu}, \citenamefont {Liu}, \citenamefont
  {Liu}, \citenamefont {Zha}, \citenamefont {Qu}, \citenamefont {Hong},
  \citenamefont {Li}, \citenamefont {Jiang}, \citenamefont {Ma}, \citenamefont
  {Hao}, \citenamefont {Zhu}, \citenamefont {Liu}, \citenamefont {Zeng},
  \citenamefont {Jayaram}, \citenamefont {Lenger}, \citenamefont {Ding},
  \citenamefont {Mo}, \citenamefont {Tanaka}, \citenamefont {Arita},
  \citenamefont {Liu}, \citenamefont {Ye}, \citenamefont {Shen}, \citenamefont
  {Wrachtrup}, \citenamefont {Huang}, \citenamefont {He}, \citenamefont {Qiao},
  \citenamefont {Liu},\ and\ \citenamefont {Liu}}]{Liu2024}%
  \BibitemOpen
  \bibfield  {author} {\bibinfo {author} {\bibfnamefont {Y.-P.}\ \bibnamefont
  {Zhu}}, \bibinfo {author} {\bibfnamefont {X.}~\bibnamefont {Chen}}, \bibinfo
  {author} {\bibfnamefont {X.-R.}\ \bibnamefont {Liu}}, \bibinfo {author}
  {\bibfnamefont {Y.}~\bibnamefont {Liu}}, \bibinfo {author} {\bibfnamefont
  {P.}~\bibnamefont {Liu}}, \bibinfo {author} {\bibfnamefont {H.}~\bibnamefont
  {Zha}}, \bibinfo {author} {\bibfnamefont {G.}~\bibnamefont {Qu}}, \bibinfo
  {author} {\bibfnamefont {C.}~\bibnamefont {Hong}}, \bibinfo {author}
  {\bibfnamefont {J.}~\bibnamefont {Li}}, \bibinfo {author} {\bibfnamefont
  {Z.}~\bibnamefont {Jiang}}, \bibinfo {author} {\bibfnamefont {X.-M.}\
  \bibnamefont {Ma}}, \bibinfo {author} {\bibfnamefont {Y.-J.}\ \bibnamefont
  {Hao}}, \bibinfo {author} {\bibfnamefont {M.-Y.}\ \bibnamefont {Zhu}},
  \bibinfo {author} {\bibfnamefont {W.}~\bibnamefont {Liu}}, \bibinfo {author}
  {\bibfnamefont {M.}~\bibnamefont {Zeng}}, \bibinfo {author} {\bibfnamefont
  {S.}~\bibnamefont {Jayaram}}, \bibinfo {author} {\bibfnamefont
  {M.}~\bibnamefont {Lenger}}, \bibinfo {author} {\bibfnamefont
  {J.}~\bibnamefont {Ding}}, \bibinfo {author} {\bibfnamefont {S.}~\bibnamefont
  {Mo}}, \bibinfo {author} {\bibfnamefont {K.}~\bibnamefont {Tanaka}}, \bibinfo
  {author} {\bibfnamefont {M.}~\bibnamefont {Arita}}, \bibinfo {author}
  {\bibfnamefont {Z.}~\bibnamefont {Liu}}, \bibinfo {author} {\bibfnamefont
  {M.}~\bibnamefont {Ye}}, \bibinfo {author} {\bibfnamefont {D.}~\bibnamefont
  {Shen}}, \bibinfo {author} {\bibfnamefont {J.}~\bibnamefont {Wrachtrup}},
  \bibinfo {author} {\bibfnamefont {Y.}~\bibnamefont {Huang}}, \bibinfo
  {author} {\bibfnamefont {R.-H.}\ \bibnamefont {He}}, \bibinfo {author}
  {\bibfnamefont {S.}~\bibnamefont {Qiao}}, \bibinfo {author} {\bibfnamefont
  {Q.}~\bibnamefont {Liu}}, \ and\ \bibinfo {author} {\bibfnamefont
  {C.}~\bibnamefont {Liu}},\ }\href {\doibase 10.1038/s41586-024-07023-w}
  {\bibfield  {journal} {\bibinfo  {journal} {Nature}\ }\textbf {\bibinfo
  {volume} {626}},\ \bibinfo {pages} {523} (\bibinfo {year}
  {2024})}\BibitemShut {NoStop}%
\bibitem [{\citenamefont {Krempask{\'y}}\ \emph {et~al.}(2024)\citenamefont
  {Krempask{\'y}}, \citenamefont {{\v S}mejkal}, \citenamefont {D'Souza},
  \citenamefont {Hajlaoui}, \citenamefont {Springholz}, \citenamefont
  {Uhl{\'\i}{\v r}ov{\'a}}, \citenamefont {Alarab}, \citenamefont
  {Constantinou}, \citenamefont {Strocov}, \citenamefont {Usanov},
  \citenamefont {Pudelko}, \citenamefont {Gonz{\'a}lez-Hern{\'a}ndez},
  \citenamefont {Birk~Hellenes}, \citenamefont {Jansa}, \citenamefont
  {Reichlov{\'a}}, \citenamefont {{\v S}ob{\'a}{\v n}}, \citenamefont
  {Gonzalez~Betancourt}, \citenamefont {Wadley}, \citenamefont {Sinova},
  \citenamefont {Kriegner}, \citenamefont {Min{\'a}r}, \citenamefont {Dil},\
  and\ \citenamefont {Jungwirth}}]{Krempasky2024}%
  \BibitemOpen
  \bibfield  {author} {\bibinfo {author} {\bibfnamefont {J.}~\bibnamefont
  {Krempask{\'y}}}, \bibinfo {author} {\bibfnamefont {L.}~\bibnamefont {{\v
  S}mejkal}}, \bibinfo {author} {\bibfnamefont {S.~W.}\ \bibnamefont
  {D'Souza}}, \bibinfo {author} {\bibfnamefont {M.}~\bibnamefont {Hajlaoui}},
  \bibinfo {author} {\bibfnamefont {G.}~\bibnamefont {Springholz}}, \bibinfo
  {author} {\bibfnamefont {K.}~\bibnamefont {Uhl{\'\i}{\v r}ov{\'a}}}, \bibinfo
  {author} {\bibfnamefont {F.}~\bibnamefont {Alarab}}, \bibinfo {author}
  {\bibfnamefont {P.~C.}\ \bibnamefont {Constantinou}}, \bibinfo {author}
  {\bibfnamefont {V.}~\bibnamefont {Strocov}}, \bibinfo {author} {\bibfnamefont
  {D.}~\bibnamefont {Usanov}}, \bibinfo {author} {\bibfnamefont {W.~R.}\
  \bibnamefont {Pudelko}}, \bibinfo {author} {\bibfnamefont {R.}~\bibnamefont
  {Gonz{\'a}lez-Hern{\'a}ndez}}, \bibinfo {author} {\bibfnamefont
  {A.}~\bibnamefont {Birk~Hellenes}}, \bibinfo {author} {\bibfnamefont
  {Z.}~\bibnamefont {Jansa}}, \bibinfo {author} {\bibfnamefont
  {H.}~\bibnamefont {Reichlov{\'a}}}, \bibinfo {author} {\bibfnamefont
  {Z.}~\bibnamefont {{\v S}ob{\'a}{\v n}}}, \bibinfo {author} {\bibfnamefont
  {R.~D.}\ \bibnamefont {Gonzalez~Betancourt}}, \bibinfo {author}
  {\bibfnamefont {P.}~\bibnamefont {Wadley}}, \bibinfo {author} {\bibfnamefont
  {J.}~\bibnamefont {Sinova}}, \bibinfo {author} {\bibfnamefont
  {D.}~\bibnamefont {Kriegner}}, \bibinfo {author} {\bibfnamefont
  {J.}~\bibnamefont {Min{\'a}r}}, \bibinfo {author} {\bibfnamefont {J.~H.}\
  \bibnamefont {Dil}}, \ and\ \bibinfo {author} {\bibfnamefont
  {T.}~\bibnamefont {Jungwirth}},\ }\href {\doibase 10.1038/s41586-023-06907-7}
  {\bibfield  {journal} {\bibinfo  {journal} {Nature}\ }\textbf {\bibinfo
  {volume} {626}},\ \bibinfo {pages} {517} (\bibinfo {year}
  {2024})}\BibitemShut {NoStop}%
\bibitem [{\citenamefont {Lee}\ \emph {et~al.}(2024)\citenamefont {Lee},
  \citenamefont {Lee}, \citenamefont {Jung}, \citenamefont {Jung},
  \citenamefont {Kim}, \citenamefont {Lee}, \citenamefont {Seok}, \citenamefont
  {Kim}, \citenamefont {Park}, \citenamefont {\ifmmode~\check{S}\else
  \v{S}\fi{}mejkal}, \citenamefont {Kang},\ and\ \citenamefont
  {Kim}}]{Lee2024}%
  \BibitemOpen
  \bibfield  {author} {\bibinfo {author} {\bibfnamefont {S.}~\bibnamefont
  {Lee}}, \bibinfo {author} {\bibfnamefont {S.}~\bibnamefont {Lee}}, \bibinfo
  {author} {\bibfnamefont {S.}~\bibnamefont {Jung}}, \bibinfo {author}
  {\bibfnamefont {J.}~\bibnamefont {Jung}}, \bibinfo {author} {\bibfnamefont
  {D.}~\bibnamefont {Kim}}, \bibinfo {author} {\bibfnamefont {Y.}~\bibnamefont
  {Lee}}, \bibinfo {author} {\bibfnamefont {B.}~\bibnamefont {Seok}}, \bibinfo
  {author} {\bibfnamefont {J.}~\bibnamefont {Kim}}, \bibinfo {author}
  {\bibfnamefont {B.~G.}\ \bibnamefont {Park}}, \bibinfo {author}
  {\bibfnamefont {L.}~\bibnamefont {\ifmmode~\check{S}\else \v{S}\fi{}mejkal}},
  \bibinfo {author} {\bibfnamefont {C.-J.}\ \bibnamefont {Kang}}, \ and\
  \bibinfo {author} {\bibfnamefont {C.}~\bibnamefont {Kim}},\ }\href {\doibase
  10.1103/PhysRevLett.132.036702} {\bibfield  {journal} {\bibinfo  {journal}
  {Phys. Rev. Lett.}\ }\textbf {\bibinfo {volume} {132}},\ \bibinfo {pages}
  {036702} (\bibinfo {year} {2024})}\BibitemShut {NoStop}%
\bibitem [{\citenamefont {Fedchenko}\ \emph {et~al.}(2024)\citenamefont
  {Fedchenko}, \citenamefont {Min\'ar}, \citenamefont {Akashdeep},
  \citenamefont {D’Souza}, \citenamefont {Vasilyev}, \citenamefont {Tkach},
  \citenamefont {Odenbreit}, \citenamefont {Nguyen}, \citenamefont
  {Kutnyakhov}, \citenamefont {Wind}, \citenamefont {Wenthaus}, \citenamefont
  {Scholz}, \citenamefont {Rossnagel}, \citenamefont {Hoesch}, \citenamefont
  {Aeschlimann}, \citenamefont {Stadtmüller}, \citenamefont {Kläui},
  \citenamefont {Schönhense}, \citenamefont {Jungwirth}, \citenamefont
  {Hellenes}, \citenamefont {Jakob}, \citenamefont {\ifmmode~\check{S}\else
  \v{S}\fi{}mejkal}, \citenamefont {Sinova},\ and\ \citenamefont
  {Elmers}}]{Fedchenko2024}%
  \BibitemOpen
  \bibfield  {author} {\bibinfo {author} {\bibfnamefont {O.}~\bibnamefont
  {Fedchenko}}, \bibinfo {author} {\bibfnamefont {J.}~\bibnamefont {Min\'ar}},
  \bibinfo {author} {\bibfnamefont {A.}~\bibnamefont {Akashdeep}}, \bibinfo
  {author} {\bibfnamefont {S.~W.}\ \bibnamefont {D’Souza}}, \bibinfo {author}
  {\bibfnamefont {D.}~\bibnamefont {Vasilyev}}, \bibinfo {author}
  {\bibfnamefont {O.}~\bibnamefont {Tkach}}, \bibinfo {author} {\bibfnamefont
  {L.}~\bibnamefont {Odenbreit}}, \bibinfo {author} {\bibfnamefont
  {Q.}~\bibnamefont {Nguyen}}, \bibinfo {author} {\bibfnamefont
  {D.}~\bibnamefont {Kutnyakhov}}, \bibinfo {author} {\bibfnamefont
  {N.}~\bibnamefont {Wind}}, \bibinfo {author} {\bibfnamefont {L.}~\bibnamefont
  {Wenthaus}}, \bibinfo {author} {\bibfnamefont {M.}~\bibnamefont {Scholz}},
  \bibinfo {author} {\bibfnamefont {K.}~\bibnamefont {Rossnagel}}, \bibinfo
  {author} {\bibfnamefont {M.}~\bibnamefont {Hoesch}}, \bibinfo {author}
  {\bibfnamefont {M.}~\bibnamefont {Aeschlimann}}, \bibinfo {author}
  {\bibfnamefont {B.}~\bibnamefont {Stadtmüller}}, \bibinfo {author}
  {\bibfnamefont {M.}~\bibnamefont {Kläui}}, \bibinfo {author} {\bibfnamefont
  {G.}~\bibnamefont {Schönhense}}, \bibinfo {author} {\bibfnamefont
  {T.}~\bibnamefont {Jungwirth}}, \bibinfo {author} {\bibfnamefont {A.~B.}\
  \bibnamefont {Hellenes}}, \bibinfo {author} {\bibfnamefont {G.}~\bibnamefont
  {Jakob}}, \bibinfo {author} {\bibfnamefont {L.}~\bibnamefont
  {\ifmmode~\check{S}\else \v{S}\fi{}mejkal}}, \bibinfo {author} {\bibfnamefont
  {J.}~\bibnamefont {Sinova}}, \ and\ \bibinfo {author} {\bibfnamefont {H.-J.}\
  \bibnamefont {Elmers}},\ }\href {\doibase 10.1126/sciadv.adj4883} {\bibfield
  {journal} {\bibinfo  {journal} {Science Advances}\ }\textbf {\bibinfo
  {volume} {10}},\ \bibinfo {pages} {eadj4883} (\bibinfo {year}
  {2024})}\BibitemShut {NoStop}%
\bibitem [{\citenamefont {Reimers}\ \emph {et~al.}(2024)\citenamefont
  {Reimers}, \citenamefont {Odenbreit}, \citenamefont {{\v S}mejkal},
  \citenamefont {Strocov}, \citenamefont {Constantinou}, \citenamefont
  {Hellenes}, \citenamefont {Jaeschke~Ubiergo}, \citenamefont {Campos},
  \citenamefont {Bharadwaj}, \citenamefont {Chakraborty}, \citenamefont
  {Denneulin}, \citenamefont {Shi}, \citenamefont {Dunin-Borkowski},
  \citenamefont {Das}, \citenamefont {Kl{\"a}ui}, \citenamefont {Sinova},\ and\
  \citenamefont {Jourdan}}]{Reimers2024}%
  \BibitemOpen
  \bibfield  {author} {\bibinfo {author} {\bibfnamefont {S.}~\bibnamefont
  {Reimers}}, \bibinfo {author} {\bibfnamefont {L.}~\bibnamefont {Odenbreit}},
  \bibinfo {author} {\bibfnamefont {L.}~\bibnamefont {{\v S}mejkal}}, \bibinfo
  {author} {\bibfnamefont {V.~N.}\ \bibnamefont {Strocov}}, \bibinfo {author}
  {\bibfnamefont {P.}~\bibnamefont {Constantinou}}, \bibinfo {author}
  {\bibfnamefont {A.~B.}\ \bibnamefont {Hellenes}}, \bibinfo {author}
  {\bibfnamefont {R.}~\bibnamefont {Jaeschke~Ubiergo}}, \bibinfo {author}
  {\bibfnamefont {W.~H.}\ \bibnamefont {Campos}}, \bibinfo {author}
  {\bibfnamefont {V.~K.}\ \bibnamefont {Bharadwaj}}, \bibinfo {author}
  {\bibfnamefont {A.}~\bibnamefont {Chakraborty}}, \bibinfo {author}
  {\bibfnamefont {T.}~\bibnamefont {Denneulin}}, \bibinfo {author}
  {\bibfnamefont {W.}~\bibnamefont {Shi}}, \bibinfo {author} {\bibfnamefont
  {R.~E.}\ \bibnamefont {Dunin-Borkowski}}, \bibinfo {author} {\bibfnamefont
  {S.}~\bibnamefont {Das}}, \bibinfo {author} {\bibfnamefont {M.}~\bibnamefont
  {Kl{\"a}ui}}, \bibinfo {author} {\bibfnamefont {J.}~\bibnamefont {Sinova}}, \
  and\ \bibinfo {author} {\bibfnamefont {M.}~\bibnamefont {Jourdan}},\ }\href
  {\doibase 10.1038/s41467-024-46476-5} {\bibfield  {journal} {\bibinfo
  {journal} {Nature Communications}\ }\textbf {\bibinfo {volume} {15}},\
  \bibinfo {pages} {2116} (\bibinfo {year} {2024})}\BibitemShut {NoStop}%
\bibitem [{\citenamefont {Ivanov}(2001)}]{Ivanov2001}%
  \BibitemOpen
  \bibfield  {author} {\bibinfo {author} {\bibfnamefont {D.~A.}\ \bibnamefont
  {Ivanov}},\ }\href {\doibase 10.1103/PhysRevLett.86.268} {\bibfield
  {journal} {\bibinfo  {journal} {Phys. Rev. Lett.}\ }\textbf {\bibinfo
  {volume} {86}},\ \bibinfo {pages} {268} (\bibinfo {year} {2001})}\BibitemShut
  {NoStop}%
\bibitem [{\citenamefont {Hansson}\ \emph {et~al.}(2004)\citenamefont
  {Hansson}, \citenamefont {Oganesyan},\ and\ \citenamefont
  {Sondhi}}]{Hansson2004}%
  \BibitemOpen
  \bibfield  {author} {\bibinfo {author} {\bibfnamefont {T.}~\bibnamefont
  {Hansson}}, \bibinfo {author} {\bibfnamefont {V.}~\bibnamefont {Oganesyan}},
  \ and\ \bibinfo {author} {\bibfnamefont {S.}~\bibnamefont {Sondhi}},\ }\href
  {\doibase https://doi.org/10.1016/j.aop.2004.05.006} {\bibfield  {journal}
  {\bibinfo  {journal} {Annals of Physics}\ }\textbf {\bibinfo {volume}
  {313}},\ \bibinfo {pages} {497} (\bibinfo {year} {2004})}\BibitemShut
  {NoStop}%
\bibitem [{\citenamefont {Naka}\ \emph {et~al.}(2019)\citenamefont {Naka},
  \citenamefont {Hayami}, \citenamefont {Kusunose}, \citenamefont {Yanagi},
  \citenamefont {Motome},\ and\ \citenamefont {Seo}}]{Naka2019}%
  \BibitemOpen
  \bibfield  {author} {\bibinfo {author} {\bibfnamefont {M.}~\bibnamefont
  {Naka}}, \bibinfo {author} {\bibfnamefont {S.}~\bibnamefont {Hayami}},
  \bibinfo {author} {\bibfnamefont {H.}~\bibnamefont {Kusunose}}, \bibinfo
  {author} {\bibfnamefont {Y.}~\bibnamefont {Yanagi}}, \bibinfo {author}
  {\bibfnamefont {Y.}~\bibnamefont {Motome}}, \ and\ \bibinfo {author}
  {\bibfnamefont {H.}~\bibnamefont {Seo}},\ }\href {\doibase
  10.1038/s41467-019-12229-y} {\bibfield  {journal} {\bibinfo  {journal}
  {Nature Communications}\ }\textbf {\bibinfo {volume} {10}},\ \bibinfo {pages}
  {4305} (\bibinfo {year} {2019})}\BibitemShut {NoStop}%
\bibitem [{\citenamefont {Yuan}\ \emph {et~al.}(2020)\citenamefont {Yuan},
  \citenamefont {Wang}, \citenamefont {Luo}, \citenamefont {Rashba},\ and\
  \citenamefont {Zunger}}]{Yuan2020}%
  \BibitemOpen
  \bibfield  {author} {\bibinfo {author} {\bibfnamefont {L.-D.}\ \bibnamefont
  {Yuan}}, \bibinfo {author} {\bibfnamefont {Z.}~\bibnamefont {Wang}}, \bibinfo
  {author} {\bibfnamefont {J.-W.}\ \bibnamefont {Luo}}, \bibinfo {author}
  {\bibfnamefont {E.~I.}\ \bibnamefont {Rashba}}, \ and\ \bibinfo {author}
  {\bibfnamefont {A.}~\bibnamefont {Zunger}},\ }\href {\doibase
  10.1103/PhysRevB.102.014422} {\bibfield  {journal} {\bibinfo  {journal}
  {Phys. Rev. B}\ }\textbf {\bibinfo {volume} {102}},\ \bibinfo {pages}
  {014422} (\bibinfo {year} {2020})}\BibitemShut {NoStop}%
\bibitem [{\citenamefont {Mazin}\ \emph {et~al.}(2021)\citenamefont {Mazin},
  \citenamefont {Koepernik}, \citenamefont {Johannes}, \citenamefont
  {González-Hernández},\ and\ \citenamefont {Šmejkal}}]{Mazin2021}%
  \BibitemOpen
  \bibfield  {author} {\bibinfo {author} {\bibfnamefont {I.~I.}\ \bibnamefont
  {Mazin}}, \bibinfo {author} {\bibfnamefont {K.}~\bibnamefont {Koepernik}},
  \bibinfo {author} {\bibfnamefont {M.~D.}\ \bibnamefont {Johannes}}, \bibinfo
  {author} {\bibfnamefont {R.}~\bibnamefont {González-Hernández}}, \ and\
  \bibinfo {author} {\bibfnamefont {L.}~\bibnamefont {Šmejkal}},\ }\href
  {\doibase 10.1073/pnas.2108924118} {\bibfield  {journal} {\bibinfo  {journal}
  {Proceedings of the National Academy of Sciences}\ }\textbf {\bibinfo
  {volume} {118}},\ \bibinfo {pages} {e2108924118} (\bibinfo {year}
  {2021})}\BibitemShut {NoStop}%
\bibitem [{\citenamefont {Bhowal}\ and\ \citenamefont
  {Spaldin}(2022)}]{Spaldin2022}%
  \BibitemOpen
  \bibfield  {author} {\bibinfo {author} {\bibfnamefont {S.}~\bibnamefont
  {Bhowal}}\ and\ \bibinfo {author} {\bibfnamefont {N.~A.}\ \bibnamefont
  {Spaldin}},\ }\href {https://arxiv.org/abs/2212.03756} {\enquote {\bibinfo
  {title} {Magnetic octupoles as the order parameter for unconventional
  antiferromagnetism},}\ } (\bibinfo {year} {2022}),\ \Eprint
  {http://arxiv.org/abs/2212.03756} {arXiv:2212.03756 [cond-mat.str-el]}
  \BibitemShut {NoStop}%
\bibitem [{\citenamefont {Guo}\ \emph {et~al.}(2023)\citenamefont {Guo},
  \citenamefont {Liu}, \citenamefont {Janson}, \citenamefont {Fulga},
  \citenamefont {{van den Brink}},\ and\ \citenamefont {Facio}}]{Guo2023}%
  \BibitemOpen
  \bibfield  {author} {\bibinfo {author} {\bibfnamefont {Y.}~\bibnamefont
  {Guo}}, \bibinfo {author} {\bibfnamefont {H.}~\bibnamefont {Liu}}, \bibinfo
  {author} {\bibfnamefont {O.}~\bibnamefont {Janson}}, \bibinfo {author}
  {\bibfnamefont {I.~C.}\ \bibnamefont {Fulga}}, \bibinfo {author}
  {\bibfnamefont {J.}~\bibnamefont {{van den Brink}}}, \ and\ \bibinfo {author}
  {\bibfnamefont {J.~I.}\ \bibnamefont {Facio}},\ }\href {\doibase
  https://doi.org/10.1016/j.mtphys.2023.100991} {\bibfield  {journal} {\bibinfo
   {journal} {Materials Today Physics}\ }\textbf {\bibinfo {volume} {32}},\
  \bibinfo {pages} {100991} (\bibinfo {year} {2023})}\BibitemShut {NoStop}%
\bibitem [{\citenamefont {Maznichenko}\ \emph {et~al.}(2024)\citenamefont
  {Maznichenko}, \citenamefont {Ernst}, \citenamefont {Maryenko}, \citenamefont
  {Dugaev}, \citenamefont {Sherman}, \citenamefont {Buczek}, \citenamefont
  {Parkin},\ and\ \citenamefont {Ostanin}}]{Ostanin2024}%
  \BibitemOpen
  \bibfield  {author} {\bibinfo {author} {\bibfnamefont {I.~V.}\ \bibnamefont
  {Maznichenko}}, \bibinfo {author} {\bibfnamefont {A.}~\bibnamefont {Ernst}},
  \bibinfo {author} {\bibfnamefont {D.}~\bibnamefont {Maryenko}}, \bibinfo
  {author} {\bibfnamefont {V.~K.}\ \bibnamefont {Dugaev}}, \bibinfo {author}
  {\bibfnamefont {E.~Y.}\ \bibnamefont {Sherman}}, \bibinfo {author}
  {\bibfnamefont {P.}~\bibnamefont {Buczek}}, \bibinfo {author} {\bibfnamefont
  {S.~S.~P.}\ \bibnamefont {Parkin}}, \ and\ \bibinfo {author} {\bibfnamefont
  {S.}~\bibnamefont {Ostanin}},\ }\href {https://arxiv.org/abs/2411.00583}
  {\enquote {\bibinfo {title} {Fragile altermagnetism and orbital disorder in
  mott insulator latio$_3$},}\ } (\bibinfo {year} {2024}),\ \Eprint
  {http://arxiv.org/abs/2411.00583} {arXiv:2411.00583 [cond-mat.mtrl-sci]}
  \BibitemShut {NoStop}%
\bibitem [{\citenamefont {Zhu}\ \emph {et~al.}(2023)\citenamefont {Zhu},
  \citenamefont {Zhuang}, \citenamefont {Wu},\ and\ \citenamefont
  {Yan}}]{Zhu2023}%
  \BibitemOpen
  \bibfield  {author} {\bibinfo {author} {\bibfnamefont {D.}~\bibnamefont
  {Zhu}}, \bibinfo {author} {\bibfnamefont {Z.-Y.}\ \bibnamefont {Zhuang}},
  \bibinfo {author} {\bibfnamefont {Z.}~\bibnamefont {Wu}}, \ and\ \bibinfo
  {author} {\bibfnamefont {Z.}~\bibnamefont {Yan}},\ }\href {\doibase
  10.1103/PhysRevB.108.184505} {\bibfield  {journal} {\bibinfo  {journal}
  {Phys. Rev. B}\ }\textbf {\bibinfo {volume} {108}},\ \bibinfo {pages}
  {184505} (\bibinfo {year} {2023})}\BibitemShut {NoStop}%
\bibitem [{\citenamefont {Heung}\ and\ \citenamefont
  {Franz}(2024)}]{Sunny2024}%
  \BibitemOpen
  \bibfield  {author} {\bibinfo {author} {\bibfnamefont {S.~T.~F.}\
  \bibnamefont {Heung}}\ and\ \bibinfo {author} {\bibfnamefont
  {M.}~\bibnamefont {Franz}},\ }\href@noop {} {\enquote {\bibinfo {title}
  {Superconducting instabilities of altermagnetic metals},}\ } (\bibinfo {year}
  {2024}),\ \Eprint {http://arxiv.org/abs/unpublished} {unpublished}
  \BibitemShut {NoStop}%
\bibitem [{\citenamefont {Vafek}\ \emph {et~al.}(2001)\citenamefont {Vafek},
  \citenamefont {Melikyan}, \citenamefont {Franz},\ and\ \citenamefont
  {Te\ifmmode \check{s}\else \v{s}\fi{}anovi\ifmmode~\acute{c}\else
  \'{c}\fi{}}}]{Vafek2001}%
  \BibitemOpen
  \bibfield  {author} {\bibinfo {author} {\bibfnamefont {O.}~\bibnamefont
  {Vafek}}, \bibinfo {author} {\bibfnamefont {A.}~\bibnamefont {Melikyan}},
  \bibinfo {author} {\bibfnamefont {M.}~\bibnamefont {Franz}}, \ and\ \bibinfo
  {author} {\bibfnamefont {Z.}~\bibnamefont {Te\ifmmode \check{s}\else
  \v{s}\fi{}anovi\ifmmode~\acute{c}\else \'{c}\fi{}}},\ }\href {\doibase
  10.1103/PhysRevB.63.134509} {\bibfield  {journal} {\bibinfo  {journal} {Phys.
  Rev. B}\ }\textbf {\bibinfo {volume} {63}},\ \bibinfo {pages} {134509}
  (\bibinfo {year} {2001})}\BibitemShut {NoStop}%
\bibitem [{\citenamefont {Liu}\ and\ \citenamefont {Franz}(2015)}]{Liu2015}%
  \BibitemOpen
  \bibfield  {author} {\bibinfo {author} {\bibfnamefont {T.}~\bibnamefont
  {Liu}}\ and\ \bibinfo {author} {\bibfnamefont {M.}~\bibnamefont {Franz}},\
  }\href {\doibase 10.1103/PhysRevB.92.134519} {\bibfield  {journal} {\bibinfo
  {journal} {Phys. Rev. B}\ }\textbf {\bibinfo {volume} {92}},\ \bibinfo
  {pages} {134519} (\bibinfo {year} {2015})}\BibitemShut {NoStop}%
\bibitem [{\citenamefont {Ben-Shach}\ \emph {et~al.}(2015)\citenamefont
  {Ben-Shach}, \citenamefont {Haim}, \citenamefont {Appelbaum}, \citenamefont
  {Oreg}, \citenamefont {Yacoby},\ and\ \citenamefont
  {Halperin}}]{Halperin2015}%
  \BibitemOpen
  \bibfield  {author} {\bibinfo {author} {\bibfnamefont {G.}~\bibnamefont
  {Ben-Shach}}, \bibinfo {author} {\bibfnamefont {A.}~\bibnamefont {Haim}},
  \bibinfo {author} {\bibfnamefont {I.}~\bibnamefont {Appelbaum}}, \bibinfo
  {author} {\bibfnamefont {Y.}~\bibnamefont {Oreg}}, \bibinfo {author}
  {\bibfnamefont {A.}~\bibnamefont {Yacoby}}, \ and\ \bibinfo {author}
  {\bibfnamefont {B.~I.}\ \bibnamefont {Halperin}},\ }\href {\doibase
  10.1103/PhysRevB.91.045403} {\bibfield  {journal} {\bibinfo  {journal} {Phys.
  Rev. B}\ }\textbf {\bibinfo {volume} {91}},\ \bibinfo {pages} {045403}
  (\bibinfo {year} {2015})}\BibitemShut {NoStop}%
\bibitem [{\citenamefont {Doll}\ and\ \citenamefont
  {N\"abauer}(1961)}]{Doll1961}%
  \BibitemOpen
  \bibfield  {author} {\bibinfo {author} {\bibfnamefont {R.}~\bibnamefont
  {Doll}}\ and\ \bibinfo {author} {\bibfnamefont {M.}~\bibnamefont
  {N\"abauer}},\ }\href {\doibase 10.1103/PhysRevLett.7.51} {\bibfield
  {journal} {\bibinfo  {journal} {Phys. Rev. Lett.}\ }\textbf {\bibinfo
  {volume} {7}},\ \bibinfo {pages} {51} (\bibinfo {year} {1961})}\BibitemShut
  {NoStop}%
\bibitem [{\citenamefont {Little}\ and\ \citenamefont
  {Parks}(1962)}]{Little1962}%
  \BibitemOpen
  \bibfield  {author} {\bibinfo {author} {\bibfnamefont {W.~A.}\ \bibnamefont
  {Little}}\ and\ \bibinfo {author} {\bibfnamefont {R.~D.}\ \bibnamefont
  {Parks}},\ }\href {\doibase 10.1103/PhysRevLett.9.9} {\bibfield  {journal}
  {\bibinfo  {journal} {Phys. Rev. Lett.}\ }\textbf {\bibinfo {volume} {9}},\
  \bibinfo {pages} {9} (\bibinfo {year} {1962})}\BibitemShut {NoStop}%
\bibitem [{\citenamefont {Lafarge}\ \emph {et~al.}(1993)\citenamefont
  {Lafarge}, \citenamefont {Joyez}, \citenamefont {Esteve}, \citenamefont
  {Urbina},\ and\ \citenamefont {Devoret}}]{Devoret1993}%
  \BibitemOpen
  \bibfield  {author} {\bibinfo {author} {\bibfnamefont {P.}~\bibnamefont
  {Lafarge}}, \bibinfo {author} {\bibfnamefont {P.}~\bibnamefont {Joyez}},
  \bibinfo {author} {\bibfnamefont {D.}~\bibnamefont {Esteve}}, \bibinfo
  {author} {\bibfnamefont {C.}~\bibnamefont {Urbina}}, \ and\ \bibinfo {author}
  {\bibfnamefont {M.~H.}\ \bibnamefont {Devoret}},\ }\href {\doibase
  10.1103/PhysRevLett.70.994} {\bibfield  {journal} {\bibinfo  {journal} {Phys.
  Rev. Lett.}\ }\textbf {\bibinfo {volume} {70}},\ \bibinfo {pages} {994}
  (\bibinfo {year} {1993})}\BibitemShut {NoStop}%
\bibitem [{\citenamefont {Mannila}\ \emph {et~al.}(2019)\citenamefont
  {Mannila}, \citenamefont {Maisi}, \citenamefont {Nguyen}, \citenamefont
  {Marcus},\ and\ \citenamefont {Pekola}}]{Pekola2019}%
  \BibitemOpen
  \bibfield  {author} {\bibinfo {author} {\bibfnamefont {E.~T.}\ \bibnamefont
  {Mannila}}, \bibinfo {author} {\bibfnamefont {V.~F.}\ \bibnamefont {Maisi}},
  \bibinfo {author} {\bibfnamefont {H.~Q.}\ \bibnamefont {Nguyen}}, \bibinfo
  {author} {\bibfnamefont {C.~M.}\ \bibnamefont {Marcus}}, \ and\ \bibinfo
  {author} {\bibfnamefont {J.~P.}\ \bibnamefont {Pekola}},\ }\href {\doibase
  10.1103/PhysRevB.100.020502} {\bibfield  {journal} {\bibinfo  {journal}
  {Phys. Rev. B}\ }\textbf {\bibinfo {volume} {100}},\ \bibinfo {pages}
  {020502} (\bibinfo {year} {2019})}\BibitemShut {NoStop}%
\bibitem [{\citenamefont {Chakraborty}\ and\ \citenamefont
  {Black-Schaffer}(2024)}]{Chakraborty2024}%
  \BibitemOpen
  \bibfield  {author} {\bibinfo {author} {\bibfnamefont {D.}~\bibnamefont
  {Chakraborty}}\ and\ \bibinfo {author} {\bibfnamefont {A.~M.}\ \bibnamefont
  {Black-Schaffer}},\ }\href {https://arxiv.org/abs/2309.14427} {\enquote
  {\bibinfo {title} {Zero-field finite-momentum and field-induced
  superconductivity in altermagnets},}\ } (\bibinfo {year} {2024}),\ \Eprint
  {http://arxiv.org/abs/2309.14427} {arXiv:2309.14427 [cond-mat.supr-con]}
  \BibitemShut {NoStop}%
\end{thebibliography}%

\end{document}